\documentstyle[12pt,aaspp4,flushrt]{article}

\newcommand{\apg}{\gtrsim}
\newcommand{\apl}{\lesssim}
\newcommand{\etal}{et al.}

\begin{document}

\lefthead{Chen \etal}
\righthead{}


\title{The Las Campanas Infrared Survey. III. The $H$-band Imaging Survey and 
the Near-Infrared and Optical Photometric Catalogs}

\author{HSIAO-WEN CHEN\altaffilmark{1}, 
P. J. McCARTHY\altaffilmark{1}, 
R. O. MARZKE\altaffilmark{1,3}
J. WILSON\altaffilmark{1},
R. G. CARLBERG\altaffilmark{2}, 
A. E. FIRTH\altaffilmark{4}, 
S. E. PERSSON\altaffilmark{1}, 
C. N. SABBEY\altaffilmark{4},
J. R. LEWIS\altaffilmark{4},
R. G. McMAHON\altaffilmark{4},
O. LAHAV\altaffilmark{4}, 
R. S. ELLIS\altaffilmark{5}, 
P. MARTINI\altaffilmark{1}, 
R. G. ABRAHAM\altaffilmark{2},
A. OEMLER\altaffilmark{1}, 
D. C. MURPHY\altaffilmark{1},
R. S. SOMERVILLE\altaffilmark{4}
M. G. BECKETT\altaffilmark{1,4},
C. D. MACKAY\altaffilmark{4}
}

\altaffiltext{1}{Carnegie Observatories, 813 Santa Barbara St, Pasadena, 
CA 91101, U.S.A.}
\altaffiltext{2}{Department of Astronomy, University of Toronto, Toronto ON, 
M5S~3H8 Canada}
\altaffiltext{3}{Department of Astronomy and Physics, San Francisco State 
University, San Francisco, CA 94132-4163, U.S.A.}
\altaffiltext{4}{Institute of Astronomy, Cambridge CB3 OHA, England, UK}
\altaffiltext{5}{Department of Astronomy, Caltech 105-24, Pasadena, CA 91125, 
U.S.A.}

\newpage

\renewcommand{\thefootnote}{\fnsymbol{footnote}}
\setcounter{footnote}{1}

\begin{abstract}

  The Las Campanas Infrared Survey, based on broad-band optical and 
near-infrared photometry, is designed to robustly identify a statistically 
significant and representative sample of evolved galaxies at redshifts $z>1$. 
We have completed an $H$-band imaging survey over 1.1 square degrees of sky in
six separate fields.  The average $5\,\sigma$ detection limit in a four 
arcsecond diameter aperture is $H\sim 20.8$.  Here we describe the design of 
the survey, the observation strategies, data reduction techniques, and object 
identification procedures.  We present sample near-infrared and optical 
photometric catalogs for objects identified in two survey fields.  The optical
images of the Hubble Deep Field South region obtained from the literature reach
$5\,\sigma$ detection thresholds in a four arcsecond diameter aperture of 
$U\sim 24.6$, $B\sim 26.1$, $V\sim 25.6$, $R\sim 25.1$, and $I\sim 24.2$ 
magnitude.  The optical images of the Chandra Deep Field South region obtained 
from our own observations reach $5\,\sigma$ detection thresholds in a four 
arcsecond diameter aperture of $V\sim 26.8$, $R\sim 26.2$, $I\sim 25.3$, and 
$z'\sim 23.7$ mag.  We perform object detection in all bandpasses and identify 
$\apg$ 54,000 galaxies over 1,408 arcmin$^2$ of sky in the two fields.  Of 
these galaxies, $\sim$ 14,000 are detected in the $H$-band and $\sim$ 2,000 
have the colors of evolved galaxies, $I - H \apg 3$, at $z\apg 1$.  We find 
that (1) the differential number counts $N(m)$ for the $H$-band detected 
objects has a slope of $d\log\,N(m)/dm = 0.45 \pm 0.01$ mag$^{-2}$ at $H\apl 
19$ and $0.27 \pm 0.01$ mag$^{-2}$ at $H \apg 19$ with a mean surface density 
$\approx 7,200$ degree$^{-2}$ mag$^{-1}$ at $H = 19$.  In addition, we find 
that (2) the differential number counts for the $H$-band detected red objects 
has a very steep slope, $d\log\,N(m; I-H\apg 3)/dm = 0.84 \pm 0.06$ mag$^{-2}$
at $H\apl 20$ and $0.32 \pm 0.07$ mag$^{-2}$ at $H\apg 20$, with a mean surface
density $\approx 3,000$ degree$^{-2}$ mag$^{-1}$ at $H=20$.  Finally, we find 
that (3) galaxies with red optical to near-IR colors ($I-H > 3$) constitute 
$\approx$ 20\% of the $H$-band detected galaxies at $H\apl 21$, but only 
$\approx$ 2\% at $H\apl 19$.  We show that red galaxies are strongly clustered,
which results in a strong field to field variation in their surface density.  
Comparisons of observations and predictions based on various formation 
scenarios indicate that these red galaxies are consistent with mildly evolving 
early-type galaxies at $z\sim 1$, although with a significant amount of 
on-going star formation as indicated by the large scatter in their $V-I$ 
colors.

\end{abstract}


\newpage

\renewcommand{\thefootnote}{\fnsymbol{footnote}}

\section{INTRODUCTION}

  Different galaxy formation scenarios provide distinct predictions for the 
space densities and masses of evolved galaxies at redshifts beyond $z = 1$.  In
monolithic collapse scenarios, massive galaxies form early over a dynamical 
time (e.g.\ Eggen, Lynden-Bell, \& Sandage 1962) and passively evolve to the 
present epoch.  The co-moving space density of evolved galaxies is therefore 
expected to remain constant and the intrinsic luminosities of these galaxies 
are expected to increase gradually with increasing redshift.  In hierarchical 
formation scenarios, massive galaxies form later through mergers of low-mass 
galaxies over a significant fraction of the Hubble time (White \& Rees 1978). 
The bulk of star formation and mass assembly occurs much later and the
co-moving space density of evolved galaxies is expected to decline steeply with
redshift at $z>1$.  While various surveys have identified some luminous galaxy
populations at $z>2$ (e.g.\ Steidel \etal\ 1999; Blain \etal\ 1999), the 
connection between these galaxies and typical present-day galaxies is unclear.
On the other hand, comparisons of statistical properties of evolved 
high-redshift galaxies and local elliptical galaxies may provide a direct means
of discriminating between competing galaxy formation scenarios (e.g.\ Kauffmann
\& Charlot 1998).

  Evolved galaxies may be characterized by their intrinsically red colors due 
to the lack of ongoing star formation that provides most of the UV light in 
typical galaxy spectral energy distributions.  At redshifts beyond unity, where
the UV region of the spectrum is redshifted to the visible, optical galaxy 
surveys are insensitive to evolved galaxies, because of their steep UV spectra.
Identifications of evolved high-redshift galaxies therefore rely on 
observations carried out at near-infrared wavelengths.  Various studies based 
on near-infrared surveys have, however, yielded inconsistent measurements of 
the space density of evolved galaxies (Kauffmann, Charlot, \& White 1996; 
Totani \& Yoshii 1997; Zepf 1997; Franceschini \etal\ 1998; Ben\'itez \etal\ 
1999; Menanteau \etal\ 1999; Barger \etal\ 1999; Schade \etal\ 1999; Broadhurst
\& Bouwens 2000; Daddi \etal\ 2000; Martini 2001a).  Because most of the 
existing deep, near-infrared surveys observe a sky area of only a few tens of 
square arc-minutes, the discrepancy is likely due to significant field-to-field
variation (e.g.\ Daddi \etal\ 2000; McCarthy \etal\ 2001a,b).  Furthermore, 
because evolved galaxies lack prominent narrow-band spectral features in the UV
spectral range,  this discrepancy is also likely due to selection biases 
(e.g.\ Totani \& Yoshii 1997).  Additional uncertainty may also arise due to 
the presence of dusty star forming galaxies that exhibit red colors resembling 
evolved high-redshift galaxies (e.g.\ Graham \& Dey 1996; Smail, Ivison, \& 
Blain 1997).

  A complete survey of evolved galaxies over a large sky area is needed to 
address these issues.  Deep, wide-field near-infrared surveys have only 
recently become feasible due to the advent of large format near-infrared 
cameras (Beckett \etal\ 1998; Elston 1998; Persson \etal\ 2001).  In addition, 
various groups have demonstrated that distant galaxies may be accurately and 
reliably identified using photometric redshift techniques that incorporate 
optical and near-infrared broad-band photometry (Connolly \etal\ 1997; Sawicki,
Lin, \& Yee 1997; Lanzetta, Fern\'{a}ndez-Soto, \& Yahil 1998; 
Fern\'{a}ndez-Soto, Lanzetta, \& Yahil 1999; Fern\'{a}ndez-Soto \etal\ 2001).
For galaxies lacking strong emission or absorption features, we may still be 
able to estimate redshifts based on the presence/absence of spectral 
discontinuities using photometric redshift techniques.  Therefore, a wide-field
optical and near-infrared imaging survey, with sufficiently accurate colors to
yield reliable photometric redshifts, offers an opportunity to identify 
statistically significant and representative samples of evolved galaxies at 
$z>1$.

  We initiated the Las Campanas Infrared (LCIR) Survey to obtain and analyze 
deep near-infrared images and complementary optical images in the $V$, $R$, 
$I$, and $z'$ bandpasses over one square degree of sky at high galactic 
latitudes (Marzke \etal\ 1999; McCarthy \etal\ 2001a,b).  The survey is 
designed to identify a large number of red galaxies at $z>1$, while securing a
uniform sample of galaxies of all types to $z\sim 2$ using broad-band optical 
and near-infrared colors.  In particular, this program utilizes one of the 
largest near-infrared cameras available (CIRSI), which produces an image of 
$13' \times 13'$ contiguous field of view in a sequence of four pointings.  The
LCIR survey may serve to bridge the gap between very wide field surveys (e.g.\ 
NOAO Deep-Wide; Jannuzi \etal\ 1998) and very deep, smaller field near-infrared
imaging surveys (e.g.\ ESO Imaging Survey, da Costa 2000; NTT Deep Field, 
Moorwood, Cuby, \& Lidman 1998).  

   The primary objectives of the program are (1) to examine the nature of the 
red galaxy population and identify evolved galaxies at redshifts $z>1$, (2) to 
study the space density and luminosity evolution of evolved galaxies at 
redshifts $z\apl 2$, and (3) to measure the spatial clustering of evolved 
high-redshift galaxies (McCarthy \etal\ 2001a,b; Firth \etal\ 2001), thereby 
inferring merging rates of these galaxies for constraining theoretical models. 
The galaxy catalog together with an accompanied photometric redshift catalog 
will also be used (4) to assess luminosity and luminosity density evolution for
galaxies of different types over the redshift range between $z=0$ to 2 (Chen 
\etal\ in preparation) and (5) to study the bright-end luminosity functions for
galaxies and QSOs at redshifts beyond $z=4.5$.  Finally, galaxies identified in
the survey will be compared with objects identified in a companion VLA survey 
(6) to study the spatial distribution of weak radio sources ($\apg 10\,\mu$Jy;
Yan \etal\ in preparation).

  To date we have completed an $H$-band imaging survey over 1.1 square degrees 
of sky to a mean $5\,\sigma$ detection limit in a four arcsecond diameter 
aperture of $H\sim 20.8$.  Here we present sample optical and near-infrared 
photometric catalogs for galaxies identified in two of our fields: the Hubble 
Deep Field South (HDFS) and Chandra Deep Field South (CDFS).  These are the two
fields for which a complete set of near-infrared and optical images are 
available.  We have identified $\sim$ 24,000 galaxies over 847 ${\rm arcmin}^2$
in the HDFS region, of which $\sim 6,720$ are detected in the $H$-band survey, 
and $\sim 30,000$ galaxies over 561 ${\rm arcmin}^2$ in the CDFS region, of 
which $\sim 7,400$ are detected in the $H$-band survey.  Among the $H$-detected
galaxies in the two regions, $\sim 2,000$ have colors that match evolved 
galaxies, $I - H \apg 3$, at redshifts $z\apg 1$.  A complete near-infrared and
optical photometric catalog of the two fields may be accessed at 
http://www.ociw.edu/lcirs/catalogs.html.  Catalogs for the other survey fields
will be released as they become available.

  We describe the survey design and observation strategies in \S\ 2 and data
reduction and image processing in \S\ 3.  The procedures for identifying 
objects in multi-bandpasses, estimates of the completeness limits of the 
catalogs, and an assessment of the reliability of the photometric measurements 
are presented in \S\ 4.  The format of the optical and near-infrared 
photometric catalogs is described in \S\ 5.  Finally, we discuss statistical 
properties of the $H$-band selected galaxies in \S\ 6.  An independent analysis
based on objects identified in the HDFS region is also presented in Firth 
\etal\ (2001).  We adopt the following cosmology: $\Omega_{\rm M}=0.3$ and
$\Omega_\Lambda=0.7$ with a dimensionless Hubble constant $h = H_0/(100 \ {\rm 
km} \ {\rm s}^{-1}\ {\rm Mpc}^{-1})$ throughout the paper.

\section{Observations}


\subsection{Definition of the Las Campanas Infrared Survey}

  The goals of the LCIR survey are to robustly identify a statistically 
significant and representative sample of evolved galaxies at redshifts $z>1$, 
as well as to secure a uniform sample of galaxies of all types at intermediate
redshifts based on broad-band optical and near-infrared colors.

  To determine the image depths necessary to reach our goals, we estimated the 
expected near-infrared brightness and optical and near-infrared colors of 
evolved high-redshift galaxies using a series of evolutionary models.  We 
considered a non-evolving model using an empirical early-type galaxy template 
presented in Coleman, Wu, \& Weedman (CWW, 1980) and two evolving models for 
galaxies of solar metallicity formed in a single burst of 1 Gyr duration at 
redshifts $z_f=5$ and $z_f=10$, respectively (Bruzual \& Charlot 1993).  We 
scaled these models to have $L_H = L_{H_*} $\footnote{We adopt $M_{H_*}=-22.9+
5\log\,h$ from De Propris \etal\ (1998).} at $z=0$ (see also Marzke \etal\ 
1999).  Figure 1 shows the predicted redshift evolution of the observed 
$H$-band magnitudes and optical and near-infrared colors of evolved galaxies 
based on the three models.  Adopting $H_0 = 70 {\rm km} \ {\rm s}^{-1}\ {\rm 
Mpc}^{-1}$, we found that in the absence of dust an evolved $L_*$ galaxy may 
have an apparent $H$-band magnitude between $H = 19.5$ and 20.3 and $I-H\sim 3$
at redshift $z\sim 1$, and an apparent $H$-band magnitude between $H=20.7$ and 
22.5 and an $I - H$ color between $\sim 4.5$ and 5 at redshift $z\sim 2$.  The 
color threshold $I-H \apg 3$ therefore defines our selection criterion for 
evolved galaxies at redshifts $z>1$, which according to Figure 1 is similar to 
selections made based on $R-K_s \apg 5$.  To identify a large number of the 
evolved galaxy population, we therefore choose a target $5\,\sigma$ survey 
depth of $H\sim 21$, more than one magnitude fainter than an $L*$ elliptical 
galaxy at $z\sim 1$.  We also require that the optical imaging survey reach 
consistent depths based on the predicted optical and near-infrared colors. The 
actual depth that we achieved varied from field-to-field and often fell 
somewhat short of our design goals.

  To determine the survey area, we first estimated the number of evolved 
galaxies that are required to obtain a significant signal in the clustering 
analysis.  At a projected co-moving correlation length $r_p = 1$ Mpc, we found 
that a sample of $\sim$ 3000 galaxies are needed to reach the $5\,\sigma$ level
of significance (Marzke \etal\ 1999).  The surface density of galaxies 
satisfying $I-H \apg 3$ is, however, very uncertain.  Different measurements 
based on $K$-band surveys have been reported by several groups, ranging from 
0.3 arcmin$^{-2}$ to 3 arcmin$^{-2}$ at $K_s < 20$ (Cowie \etal\ 1994; Elston 
\etal\ 1997; Barger \etal\ 1998).  Adopting a median surface density 1 
arcmin$^{-2}$ and a mean near-infrared color $H-K_s \sim 1$ for evolved 
galaxies at $z>1$, we concluded that one square degree of sky at a $5\,\sigma$ 
limiting magnitude $H \sim 21$ will yield $\sim 3000$ red high-redshift 
galaxies. 

    The LCIR survey may serve to bridge the gap between the on-going NOAO 
Wide-Field Survey that images 18 square degrees of sky at a shallower depth 
(Jannuzi \etal\ 1998) and the existing deep, smaller field near-infrared
imaging surveys such as the NTT Deep Field (Moorwood, Cuby, \& Lidman 1998).  
We have selected six distinct equatorial and southern fields at high galactic
latitudes distributed in right ascension.  In Table 1, we list the fields,
their J2000 coordinates, and galactic latitudes. 

\subsection{Infrared Imaging Observations}

  The $H$-band imaging survey was carried out at the du Pont 2.5~m telescope 
at Las Campanas using the Cambridge Infrared Survey Instrument (CIRSI).  This 
instrument contains a sparse mosaic of four $1024 \times 1024$ Rockwell Hawaii
HgCdTe arrays (Beckett \etal\ 1998; Persson \etal\ 2001).  At the Cassegrain 
focus of the du Pont telescope, the plate scale of the camera is 0.199 arcsec 
pixel$^{-1}$.  Some of the observations were made with a doublet field 
corrector in front of the dewar window, which produced a plate scale of 0.196 
arcsec pixel$^{-1}$.  The spacing between each array is 90\% of an array 
dimension.  Therefore, one pass of four pointings $\sim$ 192 arcsec apart in a
square pattern is needed to fill in the spaces between the arrays.  A filled 
mosaic covers $\sim 13' \times 13'$.  This defines our unit field area, 
henceforth referred to as a ``tile''.

  Observations of individual pointings were composed of nine sets of three to 
five exposures 35 to 45 s in duration with dither offsets of between 8 and 13 
arcsec in a rectangular pattern.  For a typical $H$-band sky brightness of 14.7
magnitudes, we estimated that a total exposure time of 80 minutes was needed to
reach a $5\,\sigma$ limiting magnitude of $H \sim 20.3$ in a four arcsec 
diameter aperture.  The target integration time was achieved by making three 
passes on each field.  Because it was often impossible to complete three passes
of a field within one night, the observations of any given field area were 
spread over several nights.  Standard stars from Persson \etal\ (1998) were 
observed three to five times per night.  Flat field observations were made 
using a screen hung inside the dome and were formed from differences of equal 
length exposures with the dome lamps on and off.  

  Most of our fields are composed of four contiguous tiles arranged to cover a 
roughly $25' \times 25'$ field.  The HDFS field contains seven tiles arranged 
in an H-shaped pattern to avoid several very bright stars.  The NTT Deep and 
1511 fields contain tiles that are not contiguous.  This was because the 
optical data were obtained with the sparse-array geometry Bernstein-Tyson 
Camera (BTC) on the 4-m telescope at CTIO.  The journal of $H$-band imaging 
observations is given in Table 2, in which columns (1) through (5) and column 
(8) list the field, tile number, the 2000 coordinates of the tile centers, 
total exposure time, and date of observation, respectively.

\subsection{Optical Imaging Observations}

  Optical images of the LCIR survey fields were obtained either from our own 
observations or from the literature using various large format CCDs at 
different telescopes.  The journal of optical imaging observations is given in 
Table 3, which lists the field, telescope, instrument, plate scale, field of 
view (FOV), bandpasses, total exposure time, and date of observation.  A more
detailed description of the optical imaging observations will be presented
elsewhere (Marzke \etal\ in preparation).

\section{Image Processing}

  In this section, we describe the procedures of processing the near-infrared 
$H$-band images of all six fields and summarize the qualities of the available
optical images of the HDFS and CDFS regions.  In addition, we discuss 
astrometric and photometric calibration of these images. 

\subsection{The $H$-band Images}

  The $H$-band images obtained with CIRSI were processed using two independent
reduction packages, one developed by McCarthy, Wilson and Chen at Carnegie, 
the other by Lewis, Sabbey, McMahon, and Firth at the Institute of Astronomy 
(IoA).

  The Carnegie reduction pipeline proceeds as follows.  First, loop-combined 
images were formed by taking a mean of each loop of three to five exposures 
obtained without dithering the telescope.  A $5\,\sigma$ rejection algorithm 
was applied to remove cosmic ray events, if there were more than four exposures
in a loop.  Next, the loop-combined images were divided by the normalized dome 
flats and grouped together according to their locations on the sky.  The use of
a dome rather than dark-sky flat field image allows us to separate the 
pixel-to-pixel gain variations from the additive sky fringes which are 
non-negligible for these Hawaii arrays.  The flat-field images have been 
normalized to remove the gain differences between the four detectors.  These 
are generally less than 5\%, except in the case of chip \#3 for which the gain 
correction is $\sim 20$\%.  Next, a bad pixel mask was generated for each night
using the flat-field exposures.  
Initial sky frames were constructed from medians of all frames for individual 
arrays that lay within a given pointing position.  These sky frames were 
subtracted from each flat-field corrected image for the purpose of constructing
an object mask for each image.  The sky-subtracted images were registered using
the offsets derived from the centroids of 5 to 15 stars.  All objects with peak
intensities more than $10\,\sigma$ above the sky noise were identified and 
sorted into four categories according to their isophotal sizes.  Mask frames 
were then generated for each dithered exposure accordingly.  Next, a 
sliding-median sky frame was obtained by taking the median of the preceding and
succeeding three to five images with the bad pixel and object masks applied.  
After being scaled by the ratio of the modes in the images, the median sky 
frame was subtracted from the appropriate flat-field-corrected image to remove 
the fringes due to OH emission lines from the sky that are present in the 
flat-field-corrected images.  This latter scaling (typically $\pm 1$\%) allowed
us to remove non-linear temporal variations in the sky level.  Finally, the 
flat-field-corrected and sky-subtracted images were registered and combined 
using a $5\,\sigma$ rejection criterion, a bad pixel mask, and a weight 
proportional to the inverse of the sky variance to form a final stacked image 
for each pointing position.

  The contiguous $13' \times 13'$ tiles were assembled from the 16 stacked 
images corresponding to the four arrays at four pointings of the telescope.  
The array boundaries were registered using stars in the overlap region, which 
was typically 20 to 30 arcsec in extent.  The four arrays were not perfectly 
aligned with each other and a significant rotation ($\sim 0.4^\circ$) was found
necessary to bring one of the arrays into registration with the other three.  
The final image was constructed by summing the 16 registered stacked images.  A
$5\,\sigma$ rejection criterion and a weighting determined from the sky 
variance were also applied in the vicinity of the array boundaries.  A 
$1\,\sigma$ error image was formed simultaneously for each tile through 
appropriate error propagation, assuming Poisson counting statistics.

  The IoA based reduction pipeline is similar in conceptual design, although it
is written in a somewhat different architecture and contains a few operational 
differences (Sabbey \etal\ 2001).  The offsets between dithered images were 
derived via cross-correlations in the IoA pipeline rather than from image 
centroiding.  The object masks were derived from segmentation maps produced by 
SExtractor (Bertin \& Arnout 1996) rather than using a set of five fixed object
sizes as in the Carnegie pipeline.  Finally, the assembly of the final tiles 
was completed after the application of a world coordinate system derived from 
comparison with the APM catalog.  In the Carnegie pipeline the astrometry was 
derived for the completed tiles using the USNO astrometric catalog, discussed 
in \S\ 3.3.

   The images of the first three tiles of the HDFS region, the first tile of
the NTT field, and the second tile of the 1511 field contained significant
artifacts associated with instabilities in the detector reset.  These manifest
themselves as low spatial frequency bias roll-offs that change shape and 
amplitude on short timescales.  Both the Carnegie and IoA pipelines deal with 
this low frequency signal by fitting low order polynomials to sigma-clipped 
images on a line-by-line and column-by-column basis.  In the Carnegie pipeline 
this reset correction is applied after the final sky subtraction, in the IoA 
pipeline it is applied before the flat-field and sky corrections have been 
determined.  New read-out electronics were implemented in August 2000, which
largely suppress the reset instability.

  The full width at half maximum (FWHM) of a typical point spread function 
(PSF) and the $5\,\sigma$ detection threshold in a four arcsecond diameter 
aperture of each completed tile are given in columns (6) and (7) of Table 2, 
respectively.  The PSF was measured to vary by no more than 10\% across a tile,
but the $5\sigma$ sky noise was found to vary by as much as 50\% across each 
tile.  Figure 2 shows the distributions of the FWHM of a mean PSF (left) and 
the $5\,\sigma$ detection threshold estimated based on mean sky noise in a 
four arcsecond diameter aperture (right) for each completed tile.  Figure 3 
shows an example of a completed tile in the CDFS region from the Carnegie 
pipeline.  The average PSF of this image has a FWHM of $\approx$ 0.7 arcsec.  
The magnitudes of the two objects indicated by arrows are $H=19.4$ and 21.3.

\subsection{The Optical Images}

  The optical $U$, $B$, $V$, $R$, and $I$ images of an $40' \times 40'$ area
around the HDFS region were obtained and processed by a team at the Goddard 
Space Flight Center during the HDFS campaign using the BTC (Teplitz et al. 
2001).  The spatial resolutions of the final combined images were measured to 
range from ${\rm FWHM} \approx 1.4$ arcsec in the $V$ band to $\approx 1.6$ 
arcsec in the $B$ band, and the $5\,\sigma$ detection thresholds in a 
four arcsecond diameter aperture were measured to be $U\sim 24.6$, $B\sim 
26.1$, $V\sim 25.6$, $R\sim 25.1$, and $I\sim 24.2$ magnitudes using the 
photometric zero points determined by the Goddard team.

  The optical $V$, $R$, $I$, and $z'$ images of $\approx 36' \times 36'$ around
the CDFS region were obtained using the MosaicII imager on the 4-m telescope at
CTIO in November of 1999.  Individual images were processed following the 
standard mosaic image reduction procedures for data obtained with the MosaicII 
imager, registered to a common origin, and coadded to form final combined 
images (Marzke \etal\ in preparation).  The spatial resolutions of the final 
combined images were measured to range from ${\rm FWHM}\sim 1.0$ arcsec in the
$I$ and $z'$ bands to $\approx 1.2$ arcsec in the $V$ band, and the $5\,\sigma$
detection thresholds in a four arcsecond diameter aperture were measured to be 
$V\sim 26.8$, $R\sim 26.2$, $I\sim 25.3$, and $z'\sim 23.7$ mag.

\subsection{Astrometry}

  Accurate astrometry is required for followup observations and for matching 
our catalog with observations at other wavelengths.  The optical images of the 
HDFS region were corrected for astrometric distortions with an RMS error of 
0.16 pixels, corresponding to 0.07 arcsec (Teplitz \etal\ 2001).  We adopted 
the World Coordinate System (WCS) solution stored in the image headers to 
derive coordinates for objects identified in the images.  To correct 
astrometric distortions for the MosaicII images of the CDFS region, we first 
re-sampled images of individual exposures to a Cartesian grid before 
co-addition and derived a best-fit, fourth-order polynomial for the WCS 
solution by matching stars in the U.S. Naval Observatory catalog (USNO--A2.0) 
with stars identified in the images (Marzke \etal\ in preparation).  A total of
200 stars was included to obtain the best-fit astrometric solution.  The 
goodness of fit may be characterized by an RMS residual of $\approx$ 0.2 
arcsec.

  Astrometric solutions were also derived for the near-infrared images.  We 
were able to identify on average $\sim 150$ unblended and unsaturated stars 
in each $H$-band tile.  Using a third order polynomial fit we were able to 
obtain solutions with rms residuals of between 0.25 and 0.35 arcsec.

\section{Analysis}

  In this section, we describe the procedures for object detection in 
individual bandpasses and object matching between frames taken through the
various filters.  We also examine the completeness limits of our detection 
algorithm and the reliability of photometric measurements in all bandpasses.

\subsection{Object Detection}

  To obtain a uniform sample of distant galaxies of all types, it is crucial to
identify galaxies of a wide range of optical and near-infrared colors.  To 
reach this goal, we performed object detection using SExtractor for each 
bandpass in the regions where the $H$-band imaging survey was carried out.  To 
reliably identify all the faint objects in each of the optical images, we set 
the SExtractor parameters by requiring that no detections be found in the 
negative image.  First, we set the minimum area according to the FWHM of a mean
PSF.  Next, we adjusted the detection threshold to the lowest value that is
consistent with zero detections in the negative image.  To aid in deblending 
overlapping objects, we first constructed a ``white light'' image by co-adding 
all the registered optical images that had been scaled to unit exposure time 
and filter throughput (see the next section for descriptions of image 
registration).  Next, we applied SExtractor to the white light image and 
determined object extents based on their instrumental colors.  Because the PSF
varied between different optical bandpasses by at most 10\% in each field, we 
were able to reliably deblend overlapping objects without erroneously splitting
or blending individual objects in the white light image due to a large 
variation in the PSFs of different bandpasses.  This procedure takes into 
account the intensity contrast between overlapping objects in all bandpasses 
instead of a single bandpass, allowing a more accurate object deblending for 
overlapping objects with different instrumental colors.

  We adopted a different approach to reliably identify all the faint objects in
the $H$-band images.  Because of a non-uniform noise pattern between individual
arrays across a single tile, in particular in the join regions between adjacent
array images, the zero negative detection criterion would force a significant 
underestimate of the survey depth in some areas.  An example to illustrate the 
noise pattern is given in Figure 4, which shows the associated $1\,\sigma$ 
noise image of the tile shown in Figure 3.  We identified objects well into the
noise in the $H$-band images and used our subsequent calibration of the 
completeness limits and photometric errors to set the faint magnitude limit of 
the $H$-band catalog. 

  Finally, to ensure accurate measurements of optical and near-infrared colors,
we excluded objects that lie within $\approx$ 2 arcsec of the edges of each
tile.

\subsection{Image Registration and Catalog Matching}
  
  The optical $U$, $B$, $V$, $R$, and $I$ images of the HDFS region were 
registered to a common origin by the Goddard team.  The RMS dispersion in the 
object coordinates between different bandpasses was measured to be within 
$\approx 0.2$ BTC pixels (corresponding to 0.08 arcsec).  The optical $V$, $R$,
$I$, and $z'$ images of the CDFS region were registered individually for each 
area covered by a near-infrared tile, based on a solution determined using
$\sim$ 200 common stars identified in the tile.  We applied standard routines 
to obtain the transformation solutions and to transform the images.  We were 
able to register the optical images of the CDFS tiles by a simple
transformation of low-order polynomials in the $x$ and $y$ directions without 
significantly degrading the image quality.  To assess image degradation after 
the transformation, we measured the changes in the FWHM of the PSFs in 
transformed images and found an increase of no more than 5\%.  The RMS 
dispersion in the object coordinates between different bandpasses across an 
area spanned by a tile was measured to be within $\sim 0.3$ MosaicII pixels 
(corresponding to 0.08 arcsec).  The dispersion was found to be the worst (as 
much as 0.5 MosaicII pixels or 0.13 arcsec) in the join area between adjacent 
arrays of the MosaicII imager.  To merge individual optical catalogs of each 
field, we stepped through each object identified in the white light image,
examined the presence/absence of object detections at the location in 
individual bandpasses, and recorded the SExtractor measurements for detections
and placed null values for no detections in the combined optical catalog.

  The H-band and optical catalogs were merged on the basis of astrometric
position matching, rather than by transforming the H images onto the same
pixels as the optical images.  The large differences in pixel scales and in the
PSFs of the optical and IR images result in significant image degradation if 
one transforms the H images to match the optical images.  We derived a mapping 
between the H and optical coordinates from $\sim 150$ common stars in each 
tile.  The RMS dispersion between the transformed optical coordinates and the 
$H$-band coordinates was measured to be $\approx 0.2$ CIRSI pixels 
(corresponding to 0.04 arcsec).  Next, we repeated the catalog merging 
procedure described above by examining the presence/absence of object 
detections within a radius of two CIRSI pixels to the transformed object 
location in the $H$-band image for all optically detected objects.  The 
combined catalog was updated accordingly throughout the procedure.  Finally, we
compared objects in the combined optical catalog with those in the $H$-band 
catalog, identified those that appeared only in the $H$-band catalog, and 
included them to form the final combined optical and near-infrared catalog for 
each tile.

\subsection{Survey Completeness}

  Because detection efficiencies vary significantly for objects of 
different intrinsic profiles due to the strong surface brightness selection 
biases and because the noise patterns of the $H$-band images vary across 
individual tiles, particularly at the edges of individual arrays, it is crucial
to understand the sensitivities and detection efficiencies of the LCIR survey 
in order to ensure the reliability of future analyses.  To understand the 
completeness limits of the $H$-band images (similar analyses for the optical 
images will be presented in Marzke et al. in preparation), we performed a 
series of simulations, the results of which were also adopted for testing 
the photometric accuracy described in the next section.  First, we generated 
test objects of different brightness for various PSF-convolved model surface
brightness profiles.  We adopted a Moffat (1969) profile to simulate the PSFs 
and considered (1) a point source model, (2) a de Vaucouleurs' $r^{1/4}$ model 
characterized by a half-light radius $r_e = 0.3$ arcsec\footnote{One arcsec
corresponds to a projected proper distance of 5.6 $h^{-1}$kpc at $z=1$. }, and
(3) three exponential disk models characterized by scale radii of $r_s = 0.15$,
0.3, and 0.6 arcsec, respectively.  The upper boundary of the selected disk 
scale lengths was determined according to the results of Schade \etal\ (1996) 
based on galaxies identified in the CNOC survey.  These authors found a mean 
scale length $r_s \sim$ 0.6 arcsec for field galaxies at $z\sim 0.5$.  Next, we
placed the test objects of a given magnitude at 1000 random positions in an $H$
tile and repeated the object detection procedures using the same SExtractor
parameters.  Finally, we measured the recovery rate of the test objects as a 
function of $H$-band magnitude.  

  The results of the completeness test are presented in Figure 5 for Tile 1 of 
the CDFS field.  Figure 5 shows that the 50\% completeness limit of the image 
is $\sim$ 0.6 magnitudes fainter for an exponential disk of $r_s = 0.3$ arcsec 
than for the largest disk model of $r_s = 0.6$ arcsec and is an additional 
$\sim$ 0.6 magnitudes fainter for a point source.  The large difference between
the completeness limits for objects of different intrinsic profiles makes an 
accurate incompleteness correction of the survey very challenging, because in 
practice we have no prior knowledge in the intrinsic profile of each source.  
It appears that under these circumstances correcting for the incompleteness 
based on one of these curves becomes arbitrary and therefore less accurate.

  We repeated the simulations based on empirical surface brightness profiles 
determined from the image itself to address this problem and obtain a more 
meaningful estimate of the survey completeness.  It is clear in Figure 5 that 
the completeness of an image is in practice set by the limit at which we can 
identify the majority of extended sources and that the survey for the largest 
disks in the image becomes incomplete at $H$ at least 1.5 magnitudes brighter 
than the 50\% completeness limit for point sources.  First, we adopted the
stellarity index estimated by SExtractor, which according to Bertin \& Arnouts 
(1996) is nearly unity for a point source and zero for an extended object.  We 
identified objects of stellarity $\apg$ 0.95 and with an $H$-band magnitude 
between $H = 17$ and 18, and coadded the individual images of these compact 
sources to form a high signal-to-noise empirical profile for point sources.  
Next, we repeated the simulations using the empirical profile to construct a 
completeness curve as a function of $H$ for point sources.  The results are 
shown as the thick, solid curve in Figure 5.  Next, we determined the 50\% 
completeness limit $H_{50}$ for point sources and identified extended sources 
of stellarity $\apl$ 0.5 and with an $H$-band magnitude between $H = H_{50} - 
2$ and $H_{50} - 1.5$ (to include a large enough number of objects to form a 
smooth image).  We co-added the individual images of these extended sources to 
form a high signal-to-noise empirical profile for extended sources.  Finally, 
we repeated the simulations using the empirical profile to construct a 
completeness curve as a function of $H$ for extended sources.  The results are
shown as the thick, dashed curve in Figure 5.  The 95\% completeness is 
$\approx 0.5$ magnitudes shallower for extended sources than for point sources.
The completeness curve for the empirical profile of an extended source falls 
between those of $r_s=0.15$ and 0.3 arcsec, suggesting that sources with a 
large extended disk do not constitute a large portion of the faint population.
We repeated this procedure for all $H$ tiles.

\subsection{Photometric Accuracy}

  Photometric calibrations of the optical $U$, $B$, $V$, $R$, and $I$ images 
were obtained by fitting a linear function, which included a zero-point, color,
and extinction corrections, to the Landolt standard stars (Landolt 1992) 
observed on the same nights.  Photometric calibrations of the $z'$ images were 
obtained using the Sloan Digital Sky Survey secondary standards provided in 
Gunn \etal\ (2001), who adopted the AB magnitude system.  We converted the 
best-fit zero-point to the Vega magnitude system, using $z'_{\rm AB} = z' + 
0.55$ (Fukugita, Shimasaku, \& Ichikawa, 1995), for calibrating photometry of 
our objects.  The RMS dispersion of the best-fit photometric solution was found
to be $\apl$ 0.03 magnitudes in all optical images of the CDFS field (a 
detailed discussion will be presented in Marzke \etal\ in preparation).  
Photometric calibrations of the $H$-band images were obtained by fitting a 
constant term that accounts for the zero-point correction to a number of faint 
near-infrared standard stars (Persson \etal\ 1998) observed throughout every 
night.  The RMS dispersion of the best-fit photometric solution was found to be
$\apl$ 0.02 mag.  The zero-point was found to remain stable between different 
nights with a scatter of $\apl$ 0.04 mag.

  To cross check the photometric accuracy of objects identified in the HDFS and
CDFS regions, we first compared our photometry with that from the EIS project
(da Costa \etal\ 1998) on an object-by-object basis. This clearly revealed an 
offset in the $I$-band photometry for the HDFS.  Next, we compared optical 
colors of bright, compact sources (which are likely to be stars) with the 
$UBVRI$ colors of sub-dwarfs presented in Ryan (1989).  We found that the 
stellar track agrees well with Ryan's measurements in all colors for objects in 
the CDFS region, but not for objects in the HDFS region in the $U$, $V$, and 
$I$ bandpasses.  Based on the results of stellar track comparison, we concluded 
that adjustments of $U = U_{\rm Goddard} + 0.3$, $V = V_{\rm Goddard} + 0.1$, 
and $I = I_{\rm Goddard} + 0.27$ may be needed in order to bring the stellar 
colors into agreement, where $U_{\rm Goddard}$, $V_{\rm Goddard}$, and $I_{\rm 
Goddard}$ are the photometric zero-points published by the Goddard team 
at http://hires.gsfc.nasa.gov./\~\,research/hdfs-btc/.  This also brought our
photometry very close to the EIS photometry of the same objects in the HDFS
region.

  Accurate photometry of faint galaxies is hampered by the unknown intrinsic 
surface brightness profiles of the galaxies and because of strong surface 
brightness selection biases.  Accurate total flux measurements of distant 
galaxies are crucial not only for an accurate assessment of galaxy evolution in
luminosity function (e.g.\ Dalcanton 1998) but also for comparing the observed 
number counts with theoretical predictions based on different cosmological 
models.  Various approaches to measure the total fluxes of faint galaxies have
been proposed.  For example, an ``adapted'' aperture magnitude, also known as 
a Kron magnitude, measures the integrated flux within a radius that corresponds
to the first moment of the surface brightness distribution of an object and 
corrects for the missing flux outside the adopted aperture, which was found to 
be a constant fraction of the total flux for objects of different surface 
brightness profiles (Kron 1980).  A simpler way is to measure the integrated 
flux within a fixed aperture and correct for the missing flux using a growth
curve derived from the PSF of the image. 

  To understand how well various approaches are able to recover the total flux 
for objects identified in the survey, we performed a series of photometric 
tests for objects of different brightness and intrinsic surface brightness 
profiles based on the same simulations as described in \S\ 4.3.  First, we 
generated test objects of different brightness using a series of PSF-convolved 
model surface brightness profiles.  Next, we placed the test objects at 1000 
random positions in an $H$ tile and ran SExtractor on the simulated object 
image using the same photometric parameters.  Finally, we compared various 
photometric measurements with the model fluxes.

  We present in Figure 6 comparisons of model fluxes and various photometric 
measurements for objects of different intrinsic surface brightness profiles to
the 90\% completeness limit of the image.  The left most panel shows the test
results for corrected isophotal magnitudes provided by SExtractor, the center
panel for aperture magnitudes within a four arcsecond diameter, and the right
panel for the ``best'' magnitudes estimated by SExtractor, which are either
Kron magnitudes or corrected isophotal magnitudes in cases of crowded areas.
The plotted points are the median residuals and the error bars indicate the 
16th and 84th percentiles of the residuals in 1000 realizations.  Three 
interesting points are illustrated in Figure 6.  First, the corrected isophotal
magnitudes always underestimate the total flux at faint magnitudes, 
irrespective of object intrinsic surface brightness profile.  Second, the 
four arcsecond diameter aperture photometry after a small, fixed aperture 
correction appears to deliver accurate total flux measurements for all but the 
largest disk model at all magnitudes to the 90\% completeness limits.  Third, 
the SExtractor ``best'' magnitudes appear to be able to recover the total flux 
of every profile except for the largest disks at the faintest magnitudes, 
although it also appears that a small amount of ``aperture'' correction is 
needed to bring the centroids of the residuals to zero.  Because galaxies of 
scale length greater than 0.5 arcsec, corresponding to 0.8 arcsec half-light 
radius, appear to be rare (e.g.\ Yan \etal\ 1998), we conclude that the 
corrected four arcsecond aperture magnitude provides the most accurate estimate
of the total object flux.  Although the use of a relatively large aperture 
results in reduced signal-to-noise ratios and hence a loss in depth, the loss 
in precision is offset by the improved accuracy in both the total magnitude 
measurements and the determination of colors from images with a wide range of
seeing.

\subsection{Star and Galaxy Separation}

  Stellar contamination of the final galaxy catalog from the LCIR survey is 
non-negligible, particularly in the HDFS.  An accurate separation of stars from
the galaxy catalog is therefore essential for subsequent statistical analysis. 
To identify stars in the survey fields, we can either rely on morphological 
classification by comparing object sizes and shapes with the PSF, or we can 
classify objects on the basis of their spectral shapes.  Neither approach by 
itself can robustly identify stars at all signal-to-noise levels.  

  To test the reliability of the stellarity index provided in SExtractor at 
different signal levels, we measured the rate at which a point source has a 
measured stellarity index less than a threshold value in the completeness test 
described in \S\ 4.3.  The results are shown in Figure 7.  The solid curve 
indicates the fraction of the input stars that have a stellarity index $\apl 
0.5$ as a function of $H$-band magnitude.  The dashed curve indicates the 
fraction of stars that have a stellarity index $\apl 0.8$.  The dotted line 
indicates the 50\% completeness limit ($H_{50}$) of the test image for a point 
source.  We found that while SExtractor is able to uncover point sources at
fairly bright magnitudes, it becomes unreliable at around $H_{50}- 2$ 
magnitudes (i.e.\ more than 20\% of stars would be misclassified as galaxies) 
and unstable at around $H_{50}- 1.5$ magnitudes (where objects are more or less
equally divided between compact and extended sources irrespective of their true
shapes), and therefore cannot by themselves be adopted to robustly separate 
stars from galaxies.  In addition, identifying stars based on morphological 
classification alone introduces biases against compact sources such as quasars,
narrow emission line galaxies, etc.  

  To test the reliability of comparing object colors with known stellar energy
distributions, we first included a suite of empirical stellar templates 
collected from the literature (Oppenheimer \etal\ 1998; Pickles 1998; Fan 
\etal\ 2000; Leggett \etal\ 2000) in the photometric redshift likelihood 
analysis (Chen \etal\ in preparation).  The adopted stellar templates span a 
wide range in both spectral coverage, from optical to near-infrared 
wavelengths, and stellar types, from early-type O and B stars to late-type L 
and T dwarfs.  Then we compared the results of photometric redshift analysis
with spectroscopic redshift measurements collected from the literature for 
objects in the HDFS region.  We found that while all the spectroscopically 
identified stars were accurately identified as stars based on the photometric 
redshift analysis, $\sim$ 10\% of the spectroscopically identified galaxies 
were misidentified as stars.  Therefore, we conclude that photometric redshift 
analysis with our present filter set alone cannot robustly separate galaxies 
from stars.

  To improve the accuracy of star and galaxy separation, we investigated a
method of combining both approaches.  We first examined the stellarity index 
estimated for each object by SExtractor in all bandpasses.  For objects with a 
stellarity index greater than 0.95 in one or more bandpasses, we used both
stellar and galaxy templates in the photometric redshift analysis.  For objects
with a stellarity index less than 0.95 in all bandpasses, we used only galaxy 
templates in the photometric redshift analysis.  Comparison of photometric and
spectroscopic redshifts for objects in the HDFS region demonstrates that all 11
spectroscopically identified stars may be reliably identified in this 
procedure.  

  The addition of other colors (e.g.\ $J$ \& $K$) will improve the photometric 
separation of stars and galaxies.  At present, the star and galaxy separation
may be inaccurate at faint magnitudes where the stellarity index becomes 
unreliable.  However, galaxies are likely to dominate the number counts at 
faint magnitudes (see \S\ 6.2).  Figure 8 shows the stellar fraction in the 
HDFS and CDFS regions based on the results of the star and galaxy separation.  
Closed circles indicate the mean surface density of galaxies observed in the 
two fields (see \S\ 6.2 for a detailed discussion); stars indicate the surface 
density of stellar objects in the HDFS region; and open triangles indicate the 
surface density of stellar objects in the CDFS region.  It shows that the star
counts increase gradually with the $H$ magnitude, while the galaxy counts show
a much steeper slope.  The smooth, shallower slope of the star counts suggests
that stars do not contribute a significant fraction at $H\apg 19$ in both HDFS 
and CDFS fields.

\section{The Catalogs}

  The $H$-band imaging survey presently covers 1.1 square degrees of sky to a
mean $5\,\sigma$ detection limit in a four arcsecond diameter aperture of $H
\sim 20.8$.  Here we present sample near-infrared and optical photometric 
measurements for objects identified in the HDFS and CDFS regions.  We have 
performed object detection in all bandpasses and identified $\sim 24,000$ 
galaxies over 847 arcmin$^2$ in the HDFS region, of which $\sim 6,700$ are 
detected in the $H$-band survey, and $\sim 30,000$ galaxies over 561 ${\rm 
arcmin}^2$ in the CDFS region, of which $\sim 7,400$ are detected in the 
$H$-band survey.  Among the $H$-detected galaxies in the two regions, $\sim 
2,000$ have $I - H > 3$ and are likely to be primarily evolved galaxies at 
redshifts $z>1$. 

  The catalogs for each field is organized as follows:  For each object we give
an identification number, x and y pixel coordinates in the optical images, 
$\alpha$ (J2000) \& $\delta$ (J2000), the Kron radius (in arcseconds) for each 
filter, the stellarity index determined by SExtractor, $2^{''}$ and $4^{''}$ 
diameter aperture magnitudes and their associated uncertainties, the corrected 
isophotal magnitude, auto magnitude and the best magnitude and their assocaited
uncertainties as provided by SExtractor for all the objects detected in each 
filter.  We present an example of the object catalog in Table 4 and an example 
of the photometric catalog in Table 5, which list the first 50 objects 
identified in the CDFS region.  Complete near-infrared and optical photometric 
catalogs of the two fields may be accessed at 
http://www.ociw.edu/lcirs/catalogs.html.  Catalogs for the remaining fields and
their completeness limits will be posted at this address as they become
available.

\section{Discussion}

  As we identify objects in each optical and near-infrared bandpass the 
catalogs contain objects with a wide range of colors, ranging from the reddest 
objects, whether they be old or dusty, to the most actively star forming 
systems.  The full galaxy sample is therefore representative of galaxies of a 
wide range of stellar properties and redshifts, and may be used to address a 
variety of cosmological issues.  Here we examine photometric properties of
the $H$-band selected objects and derive the number-magnitude relation for the 
$H$-band selected galaxies as well as for red galaxies of different optical and
near-infrared colors.  We compare our measurements with various model 
predictions and discuss the implications.

\subsection{Distributions of Optical and Near-infrared Colors}

  We present the $I-H$ versus $H$-band color--magnitude diagram for the $\sim 
14,000$ $H$-band selected galaxies identified in the HDFS and CDFS regions in
Figure 9.  Objects classified as stars are not included.  We show the 95\% 
completeness limit of the $H$-band data in the short dashed line and the 
$5\,\sigma$ detection threshold over a four arcsecond diameter aperture
for the $I$-band images of the HDFS and CDFS regions in the dot-dashed lines.
We also present the expected color--magnitude evolution of an $L_*$ elliptical 
galaxy from $z = 2$ to the present time under different evolution scenarios for
$h=0.7$.  The curves trace the trajectories of a non-evolving elliptical galaxy
spectrum (blue), a passively evolving galaxy formed in a single burst at $z = 
10$ (red), and a galaxy formed at $z = 30$ with an exponentially declining star
formation rate (SFR $\sim \exp (-t/\tau)$) for $\tau = 1$ Gyr (green).  The 
blue dashed curve corresponds to a non-evolving $3\,L_*$ early-type galaxy.  

  Comparisons of measurements and model predictions show that the $I-H$ colors
of galaxies with $H \apl 18$ agree well with the expected colors at low 
redshifts.  In addition, they show that the median color becomes progressively 
bluer at fainter magnitudes, accompanied by a tail of galaxies with very red 
colors ($I-H\apg 3$).  These red galaxies constitute a negligible fraction of 
the total population at $H \sim 18$, but become a significant constituent by 
$H \sim 20$.  Similar trends are seen in $B-K$, $R-K$ and $I-K$ 
color--magnitude diagrams based on smaller area surveys (Elston \etal\ 1988; 
Cowie \etal\ 1994; Ellis 1997; Cowie \etal\ 1997, Menanteau \etal\ 1999).  
Furthermore, the good agreements between observed colors and model predictions 
suggest that these red galaxies are mildly evolving early-type galaxies at $z 
\sim 1$ sampled over the bright end of the luminosity function ($\sim 1 - 
5\,L_*$).  The red sequence observed in clusters with $z \sim 0.8 - 1.2$ 
populate a similar range in the $I-H$ vs. $H$ color--magnitude diagram, 
supporting the idea that the bulk of the red galaxies are evolved early-type 
galaxies at intermediate redshifts (e.g.\ Stanford \etal\ 1997, 1998; Chapman,
McCarthy, \& Persson 2000).

  We present the $V-I$ vs. $I-H$ color--color distribution of the $H$-band 
detected objects in the HDFS and CDFS regions in Figure 10.  Stars that were
classified according to the criteria described in \S\ 4.5, are indicated by
crosses and galaxies are indicated by filled circles.  Open circles indicate
galaxies that are not detected in $V$.  The solid curves trace the trajectories
of different CWW spectral templates at redshifts from $z=2$ to the present 
time: elliptical/S0 (yellow), Sab (green), Scd (cyan), and irregular (blue).  
The dotted curve shows the trajectory of a passively evolving galaxy formed in 
a single burst at redshift $z_f=10$.  The short and long dashed curves trace 
the colors of galaxies formed at $z_f=30$ with exponentially declining star 
formation rates for $\tau = 1$ and 2 Gyr, respectively.  The dash-dotted line 
corresponds to a galaxy formed at $z_f=30$ with a constant star formation rate.
Filled circles along each curve indicate the corresponding redshifts, starting 
from $z = 2$ in the lower right corner, evolving toward lower redshifts in 
steps of $\Delta z = 0.5$, and ending at $z=0$ in the lower left corner.

  The general agreement between distributions of various model predictions and 
measurements of galaxies in our catalog, and the clear separation of the 
stellar sequence suggest that most stars have been accurately identified in 
these two survey fields.  A similar separation is made in a $B-I$ vs. $I-K$ 
color--color diagram (e.g.\ Gardner 1992; Huang \etal\ 1997).  The large 
scatter in the $V-I$ colors spanned by the galaxies reveals a wide range of 
star formation histories for which our simple models provide only an idealized 
description.  In particular, the large scatter in the $V-I$ colors for galaxies
of $I-H$ colors indicative of redshifts $z\sim 1$ suggest a significant amount 
of ongoing star formation in these galaxies.  Similar results are found in 
detailed studies of morphologically selected early-type galaxies at 
intermediate redshifts that reveal a deficit of distant ellipticals with 
optical and UV colors that match with expectations based on passively evolving 
galaxies formed at high redshifts (e.g.\ Menanteau \etal\ 1999; Menanteau, 
Abraham, \& Ellis 2001).  In addition, the small number ($<\,15$\%) of galaxies
with $I-H\apg 3$ and $V-I\apg 3$ suggests that there are few objects that may 
be adequately characterized as pure passively evolving systems formed at high 
redshifts.  The trajectories of the evolving galaxy models in the color--color 
diagram suggests that these red galaxies span a significant range in redshift 
and that those of bluer $V-I$ colors largely arise at redshifts $z\apg 1.5$.  
This is supported by their angular clustering amplitude (McCarthy \etal\ 2001b)
and photometric redshift measurements (Chen \etal\ in preparation).

\subsection{The $H$-band Number Counts}

  Galaxy number counts provide a diagnostic for distinguishing between 
different evolutionary scenarios for galaxies of different colors (e.g.\ 
Metcalfe \etal\ 1996).  Comparisons of number count measurements from different
surveys also allow us to examine the accuracy of our photometric calibration.  
Figure 11 shows the $H$-band differential galaxy counts $N(m)$ derived from 
$\sim 14,000$ galaxies selected in the $H$-band survey toward the HDFS and CDFS
regions (filled circles).  We applied incompleteness corrections down to the 
95\% completeness level using the incompleteness curves determined for compact 
and extended sources in each tile.  According to the procedures described in 
\S\ 4.3, we adopted the stellarity index determined in SExtractor for each 
object and corrected for the incompleteness at faint magnitudes using the 
compact source incompleteness curve for objects of stellarity index greater 
than 0.5 and using the extended source incompleteness curve otherwise.  The 
vertical bars indicate the uncertainties in our measurements estimated using a 
variation of the ``bootstrap'' resampling technique.  The bootstrap resampling 
technique explicitly accounts for the sampling error and for photometric 
uncertainties.  We first resampled objects from the parent catalog (allowing 
for the possibility of duplication) and added random noise (within the 
photometric uncertainties) to the photometry.  Then we estimated the $H$-band 
differential number counts for the resampled, perturbed catalog.  Finally, we 
repeated the procedures a thousand times and determined the $1\,\sigma$ 
uncertainties based on the 16th and 84th percentiles of the distributions in 
each magnitude bin.

  For comparison, we have also included in Figure 11 previous $H$-band 
measurements reported by various groups.  Our measurements span six magnitudes 
in $H$ and nicely bridge the gap between the existing measurements at the 
bright (open squares from Martini 2001b)\footnote{Part of the apparent excess 
in the number count measurements by Martini (2001b) is due to an additional 
aperture correction that the author applied in the galaxy photometry to account
for missing light in the wings of galaxy profiles, which was found to be 
between 0.1 and 0.2 mag.} and faint ends (Yan \etal\ 1998).  The excellent 
agreements in both the slopes and the normalization demonstrate not only the 
accuracy of our $H$-band photometry, but also the accuracy in our star and 
galaxy separation.  The best fitting power-law slopes, $d\log\,N(m)/dm$, were 
found to be $0.45 \pm 0.01$ mag$^{-2}$ at $H\apl 19$ and $0.27 \pm 0.01$ 
mag$^{-2}$ at $H\apg 19$ with a mean surface density of $\approx 7,200$ 
degree$^{-2}$ mag$^{-1}$ at $H = 19$\footnote{We note that the uncertainties in
the number count measurements due to random errors are small given the large 
sample from our survey, but sysmatic errors are likely to dominate and more 
difficult to measure.  In particular, number count measurements at the faintest
magnitude bins are more likely to be overestimated due to photometric 
undertainties of the even fainter and more abundant populations.  As a result, 
the faint-end slope is likely to be overestimated.}.  The slope of the 
faint-end $H$-band counts is in line with that of the faint-end $K$-band counts
(Cowie \etal\ 1994; Djorgovski \etal\ 1995; Bershady, Lowenthal, \& Koo 1998; 
Metcalfe \etal\ 1996; McCracken \etal\ 2000).  The slope of the bright-end 
$H$-band counts ($H \apl 19$) is slightly shallower than that of the bright-end
$K$-band counts reported by Huang \etal\ (1997) and Metcalfe \etal\ (1996), 
although we note that our counts are not well determined at $H<16$.  

  To compare with simple model predictions, we considered a no-evolution model 
based on an early-type galaxy template (the solid curve), a passive evolution 
scenario for galaxies formed in a single burst at $z_f=10$ (the dotted curve), 
an exponentially declining star formation for galaxies formed at $z_f = 30$ 
(the short dashed curve), and a constant star formation for galaxies formed at 
$z_f = 30$ (the long dashed curve).  We adopted the $K$-band luminosity 
function derived by Gardner \etal\ (1997) and applied an $H-K\approx 0.22$ 
color correction (Mobasher, Ellis, \& Sharples 1986) to obtain an estimate for 
$M_{H_*}$.  This is appropriate for galaxies of different type because the 
$k$-correction is nearly independent of spectral type at near-infrared 
wavelengths.  We find that all but the passive evolution scenario (the dotted 
curve) agree well with the measurements.  

\subsection{Surface Density of Red Galaxies}

  In this section, we present surface density estimates of moderately red 
galaxies of $I-H\apg 3$ and compare the measurements with those of extremely 
red objects (EROs) identified in the survey.  EROs are typically selected based
on their extreme optical and near-infrared colors $R-K \apg 6$ (e.g.\ Elston, 
Rieke, \& Rieke 1988; McCarthy, Persson, \& West 1992; Hu \& Ridgway 1994).  
Here we selected EROs using $R-H\apg 5$, which according to the simulations 
presented in Figure 1 is equivalent to $R-K \apg 6$ for high-redshift evolved 
populations.

  Figure 12 shows the $H$-band differential number counts for galaxies of $I-H
\apg 3$ (filled circles) and for galaxies of $R-H \apg 5$ (squares) identified 
in the HDFS and CDFS regions.  Incompleteness corrections were applied at faint
magnitudes and the uncertainties indicated by the vertical bars were estimated 
following the same procedures described in \S\ 6.2.  The slope of the red 
counts is quite steep, $d\log\,N(m)/dm = 0.84 \pm 0.06$ mag$^{-2}$ for 
galaxies of $I-H \apg 3$ and $0.89 \pm 0.24$ mag$^{-2}$ for galaxies of $R-H 
\apg 5$ at $H \apl 20$.  The counts appear to have shallower faint-end slopes 
for both of the red subsamples---between 0.17 and 0.32 mag$^{-2}$---at $H\apg 
20$.  While both subsamples show consistent slopes at the bright end, they have
a more steeply rising $N(m)$ with the $H$ magnitude than the total $H$-band 
detected population.  The steeper bright-end slope of the red galaxies 
indicates that most of the foreground ($z\apl 1$) galaxies have been 
effectively excluded from the red samples and that the shape of the number 
counts reflects the shape of the bright-end luminosity function of the red 
population.  In addition, the consistent slopes between the two red subsamples 
suggest that the underlying luminosity function is very similar.

  For comparison, we also calculated the predicted differential number counts 
for red galaxies with $I-H \apg 3$ based on the no-evolution scenario and the 
exponentially declining star formation scenario with $\tau = 1$ Gyr that best 
fit the observed total $H$-band number counts, as discussed in \S\ 6.2.  To 
model the intrinsic luminosity function of evolved galaxies, we adopted the
population ratio of elliptical galaxies in the local universe (e.g.\ Binggeli,
Sandage, \& Tammann 1988).  We first scaled the $K$-band luminosity function 
obtained by Gardner \etal\ (1997) for the total galaxy population by $M_*(red) 
= M_*(tot) - 0.2$ and $\phi_*(red) = 0.2\,\phi_*(tot)$ and then assumed a 
faint-end slope of $\alpha = 1$.  The curves indicate the results of the 
calculations and are the same as those in Figure 11.  It appears that both 
scenarios fit the observations fairly well, but that the exponentially 
declining star formation model predicts a sharp turnover in the differential 
number counts at $H\apg 21$.  The turnover is due to the fact that at these 
faint magnitudes the evolved population is dominated by those at redshifts 
$z\apg 2$ where they would have bluer $I-H$ colors under the exponentially 
declining star formation scenario and would therefore be excluded based on the
$I-H\apg 3$ selection criterion (see Figure 9).  Because of large uncertainties
in the optical and $H$-band photometry for galaxies at these faint magnitudes, 
we cannot rule out either of the scenarios.  The moderately good agreements 
between observations and model predictions, together with the large scatter in 
the optical $V-I$ color discussed in \S\ 6.1, however suggest that there is 
still some amount of star formation going on in the red population at redshifts
$z > 1$.

  In Figure 13 we plot the cumulative surface density measurements of galaxies 
with $I-H \apg 3$ (filled circles) and of $R-H \apg 5$ (squares) as a function 
of $H$-band magnitude, in comparison with those of the total $H$-band selected 
galaxies and with previous measurements of EROs presented by Daddi \etal\ 
(2000; green squares), Yan \etal\ (2000; blue triangles), and Thompson \etal\ 
(1999; red cross).  Our surface density measurements of EROs span two 
magnitudes in $H$ and are consistent with those from previous surveys.  In 
addition, we find that red galaxies with $I-H \apg 3$ constitute $\approx$ 20\%
of the $H$-band detected galaxies at $H\apl 21$, but only $\approx$ 2\% at 
$H\apl 19$.  Specifically, we expect to observe on average $\sim 40$ galaxies 
of $I-H \apg 3$ and $\sim 8$ galaxies of $R-H \apg 5$ among $\sim$ 270 $H$-band
detected galaxies at $H\apl 21$, but only $\sim 1$ galaxy of $I-H \apg 3$ and 
less than one of $R-H \apg 5$ among $\sim 46$ $H$-band detected galaxies at $H 
\apl 19$ in a random 25-arcmin$^{2}$ region.

  On the other hand, because EROs and these moderately red galaxies are found
to exhibit strong angular clustering (Daddi \etal\ 2000; McCarthy \etal\ 
2001b; Firth \etal\ 2001), a large disperson in the surface density 
measurements is expected.  To examine the accuracy of the surface density 
measurements presented in Figure 13, we estimated the field to field variation 
by measuring the surface density of galaxies of $I-H \apg 3$ in a randomly 
selected $5\times 5$ arcmin$^2$ region in each tile and compared the 
measurements between different tiles.  We found that the total number of red 
galaxies of $H \apl 20.5$ varies from $\approx 15$ to $\approx$ 40 with a mean 
surface density of $\approx 1$ arcmin$^{-2}$ and an rms dispersion of $\approx 
0.4$ arcmin$^{-2}$.  The mean surface density estimated based on the randomly 
selected smaller areas is consistent with the measurements presented in Figure 
12 and the red counts over $0.62$ square degrees presented in McCarthy \etal\ 
(2001a,b), indicating that our surface density measurements are not 
significantly affected by the cosmic variance due to clustering.  In addition, 
following Roche \etal\ (1999), we estimate an angular clustering amplitude of 
these red galaxies to be $A \approx 4$ at $1''$ (corresponding to a correlation
angle of $\theta_0\approx 5$ arcsec), which indeed agrees well with the results
of an angular clustering analysis presented in McCarthy \etal\ (2001b).  We 
therefore conclude that a deep, wide-field survey is necessary not only to 
identify a large number of red galaxies, but also to obtain accurate estimates 
of their statistical properties.

\section{Summary}

  We present initial results from the Las Campanas Infrared Survey, which is a
wide-field near-infrared and optical imaging survey designed to reliably
identify a statistically significant and representative sample of evolved 
galaxies at redshifts $z>1$ and to secure a uniform sample of galaxies of all 
types at intermediate redshifts.  We present sample near-infrared and optical 
photometric catalogs for objects identified in two of the six survey fields, 
surrounding the Hubble Deep Field South and Chandra Deep Field South regions. 
We have identified $\sim$ 14,000 $H$-band detected galaxies over 1,408 
arcmin$^2$ of sky, $\sim$ 2,000 of which are likely to be evolved galaxies at 
$z\apg 1$ with $I - H \apg 3$.  Furthermore, we find that:

  1. Red galaxies with $I-H\apg 3$ are a significant constituent of galaxies 
with $H \apg 20$ and are consistent with mildly evolving early-type galaxies at
$z\sim 1$.  But there exists a large scatter in the $V-I$ colors spanned by the
red galaxies, indicating a significant amount of on-going star formation in 
these galaxies.  

  2. The differential number counts $N(m)$ for the $H$-band detected objects 
has a slope of $d\log\,N(m)/dm = 0.45 \pm 0.01$ mag$^{-2}$ at $H\apl 19$ and 
$0.27 \pm 0.01$ mag$^{-2}$ at $H\apg 19$ with a mean surface density $\approx 
7,200$ degree$^{-2}$ mag$^{-1}$ at $H = 19$. 

  3. The differential number counts for red objects ($I-H\apg 3$) detected in 
the $H$-band has a very steep slope, $d\log\,N(m; I-H\apg 3)/dm = 0.84 \pm 
0.06$ mag$^{-2}$ at $H\apl 20$ and $0.32 \pm 0.07$ mag$^{-2}$ at $H\apg 20$, 
with a mean surface density $\approx 3,000$ degree$^{-2}$ mag$^{-1}$ at $H=20$.

  4. Red galaxies ($I-H\apg 3$) constitute $\approx$ 20\% of the $H$-band 
detected galaxies at $H\apl 21$, but only $\approx$ 2\% at $H\apl 19$.  This
suggests that foreground galaxies ($z\apl 1$) may be effectively excluded from 
the total $H$-band population by apply the $I-H$ color selection criterion and 
that the steep bright-end slope of red galaxy number counts reflects the shape 
of the bright-end luminosity function of the red population.

  5. Red galaxies are strongly clustered, producing strong field to field 
variation in their surface density.  We estimate a mean surface density of 
$\approx 1$ arcmin$^{-2}$ with an rms dispersion of $\approx 0.4$ arcmin$^{-2}$
over $5\times 5$ arcmin$^2$ sky regions for galaxies with $H\apl 20.5$ and 
$I-H\apg 3$, implying an angular clustering amplitude of $A \approx 4$ at $1''$
and a correlation angle of $\theta_0\approx 5$ arcsec.

\acknowledgments

  We are grateful to Xiaohui Fan, Sandy Leggett, and Ben Oppenheimer for 
providing various digitized stellar templates of M dwarfs, L dwarfs, and T 
dwarfs, and to Ken Lanzetta for providing galaxy templates for the photometric 
redshift analysis.  We appreciate the expert assistance from the staffs of the 
Cerro Tololo Inter-American Observatory and Las Campanas Observatory.  This 
research was supported by the National Science Foundation under grant 
AST-9900806.  The CIRSI camera was made possible by the generous support of the
Raymond and Beverly Sackler Foundation.

\newpage

\begin{small}
\begin{deluxetable}{p{1.0in}ccc}
\tablecaption{LCIR Survey Fields}
\tablewidth{0pt}
\tablehead{\multicolumn{1}{c}{Field} & \colhead{RA (2000)} & 
\colhead{DEC (2000)} & \colhead{$|b|$}}
\startdata
NOAO-Cetus    \dotfill& 02$^{\rm h}$ 10$^{\rm m}$& $-$04$^{\circ}$ 37$'$ & 
60$^{\circ}$ \nl
CDFS     \dotfill& 03$^{\rm h}$ 32$^{\rm m}$& $-$27$^{\circ}$ 38$'$ & 
54$^{\circ}$ \nl
NTT Deep \dotfill& 12$^{\rm h}$ 05$^{\rm m}$& $-$07$^{\circ}$ 27$'$ & 
52$^{\circ}$ \nl
1511     \dotfill& 15$^{\rm h}$ 24$^{\rm m}$& $+$00$^{\circ}$ 10$'$ & 
44$^{\circ}$ \nl
SSA22     \dotfill& 22$^{\rm h}$ 00$^{\rm m}$& $+$00$^{\circ}$ 21$'$ & 
40$^{\circ}$ \nl
HDFS     \dotfill& 22$^{\rm h}$ 33$^{\rm m}$& $-$60$^{\circ}$ 39$'$ & 
49$^{\circ}$ 
\enddata
\end{deluxetable}
\end{small}

\newpage 

\small

\begin{deluxetable}{p{1.0in}ccccccr}
\tablecaption{Journal of $H$-band Imaging Observations}
\tablewidth{0pt}
\tablehead{ & & & & \colhead{Exposure} & \colhead{FWHM} & & \\
\colhead{Field} & \colhead{Tile} & \colhead{RA (2000)} & \colhead{DEC (2000)} &
 (sec) & (arcsec) & \colhead{$H_{5\,\sigma}$\tablenotemark{a}} & 
\colhead{Date} \\
\colhead{(1)} & \colhead{(2)} & \colhead{(3)} & \colhead{(4)} & \colhead{(5)} &
\colhead{(6)} & \colhead{(7)} & \colhead{(8)}}
\startdata
NOAO-Cetus \dotfill & 1 & 02:10:13.0 & $-$04:33:00.0 & 4800  & 0.80 & 20.1 & Oct 16--19 1999      \nl
                  & 2 & 02:09:22.0 & $-$04:33:00.0 & 2430  & 0.70 & 20.2 & Oct 09--09 2000            \nl
                  & 3 & 02:10:13.0 & $-$04:45:48.0 & 3300  & 0.72 & 20.2 & Oct 11--12 2000      \nl
                  & 4 & 02:09:22.0 & $-$04:45:48.0 & 2320  & 0.68 & 20.3 & Oct 09--09 2000          \nl          
CDFS \dotfill     & 1 & 03:32:36.0 & $-$27:46:30.0 & 4420  & 0.66 & 21.3 & Oct 07--09  2000       \nl
                  & 2 & 03:32:36.0 & $-$27:33:53.0 & 4860  & 0.70 & 21.2 & Oct 05--07 2000        \nl
                  & 3 & 03:31:39.0 & $-$27:33:53.0 & 5535  & 0.69 & 21.2 & Sep 15--16  2000   \nl
                  & 4 & 03:31:39.0 & $-$27:46:30.0 & 6243  & 0.80 & 20.8 & Dec 20--31 1999     \nl
NTT Deep \dotfill & 1 & 12:05:57.8 & $-$07:41:58.0 & 4700  & 0.80 & 20.5 & Feb 13--16 1999        \nl
                  & 2 & 12:05:57.8 & $-$07:21:52.0 & 4860  & 0.90 & 20.9 & Jan 01--17 2000    \nl
                  & 3 & 12:04:36.6 & $-$07:41:58.0 & 4691  & 0.80 & 20.6 & Feb 13--16 2000        \nl
                  & 4 & 12:04:36.6 & $-$07:21:52.0 & 5264  & 0.75 & 21.3 & Dec 26--30  1999         \nl
1511 \dotfill     & 2 & 15:24:07.7 & $+$00:20:16.0 & 3600  & 0.68 & 20.1 & Feb 13--16 2000          \nl
SA22 \dotfill     & 1 & 22:18:00.0 & $+$00:14:34.0 & 4321  & 0.66 & 20.1 & Oct 16--24 1999      \nl
                  & 2 & 22:17:08.0 & $+$00:14:34.0 & 3510  & 0.60 & 20.8 & Oct 06--07 2000        \nl
                  & 3 & 22:18:00.0 & $+$00:27:30.0 & 4620  & 0.77 & 20.8 & Oct 04--06 2000        \nl
                  & 4 & 22:17:08.0 & $+$00:27:30.0 & 3731  & 0.65 & 20.5 & Oct 12--13 2000      \nl
HDFS \dotfill     & 1 & 22:33:13.0 & $-$60:39:27.0 & 4826  & 0.92 & 20.5 & Oct 18--19 1999      \nl
                  & 2 & 22:34:56.0 & $-$60:39:27.0 & 3645  & 0.90 & 20.5 & Oct 20--29 1999      \nl
                  & 3 & 22:31:30.0 & $-$60:39:27.0 & 4758  & 1.06 & 20.6 & Oct 25--29 1999      \nl
                  & 4 & 22:34:56.0 & $-$60:26:50.0 & 3645  & 0.84 & 21.3 & Sep 07--07 2000        \nl
                  & 5 & 22:34:56.0 & $-$60:49:50.0 & 4230  & 1.00 & 21.1 & Sep 13--15 2000  \nl
                  & 6 & 22:31:30.0 & $-$60:49:50.0 & 6986  & 0.85 & 21.2 & Oct 04--11 2000     \nl
                  & 7 & 22:31:30.0 & $-$60:26:00.0 & 4890  & 0.85 & 21.3 & Oct 10--12 2000  
\tablenotetext{a}{The $5\,\sigma$ detection threshold was determined over a 
four arcsec diameter aperture using a mean sky noise measured in each tile.  
The actual depth may vary, depending on the PSF, in each tile.}
\enddata
\end{deluxetable}

\normalsize

\newpage 

\begin{small}
\begin{deluxetable}{p{0.95in}lrccr}
\tablecaption{Summary of Optical Imaging Data}
\tablewidth{0pt}
\tablehead{ & & & \colhead{Plate Scale} & \colhead{FOV} &  \\
\multicolumn{1}{c}{Field} & \multicolumn{1}{c}{Telescope} & 
\multicolumn{1}{c}{Detector} & \colhead{(arcsec/pixel)} & \colhead{arcmin$^2$} 
& \multicolumn{1}{c}{Bandpasses} }
\startdata
NOAO-Cetus \dotfill & CTIO/4~m & MosaicII & 0.27 & 36$\times$36 & $V$ \nl
                    & CFHT     & CFH12k   & 0.20 & 40$\times$40 & $R,I,z'$\nl
CDFS     \dotfill& CTIO/4~m & MosaicII & 0.27 & 36$\times$36 & $B,V,R,I,z'$ \nl
NTT Deep \dotfill& CTIO/4~m & BTC      & 0.43 & 29$\times$29 & $V,R,I$ \nl
                 & CTIO/4~m & MosaicII & 0.27 & 36$\times$36 & $B,z'$ \nl
1511     \dotfill& CTIO/4~m & BTC      & 0.43 & 29$\times$29 & $V,R,I$ \nl
                 & CTIO/4~m & MosaicII & 0.27 & 36$\times$36 & $B,z'$ \nl
SA22     \dotfill& CTIO/4~m & Mosaic   & 0.26 & 36$\times$36 & $B,V,R,z'$ \nl
                 & CFHT     & CFH12k   & 0.20 & 40$\times$40 & $I$ \nl
HDFS     \dotfill& CTIO/4~m & BTC      & 0.43 & 29$\times$29 & $U,B,V,R,I$\tablenotemark{b} \nl 
	         & CTIO/4~m & MosaicII & 0.27 & 36$\times$36 & $z'$ 
\tablenotetext{b}{Teplitz \etal\ (2001).}
\enddata
\end{deluxetable}
\end{small}

\newpage

\figcaption{The predicted $H$-band magnitude (left panel) and optical and
near-infrared colors (right panel) of an $L*$ elliptical galaxy as a 
function of redshift.  The solid curves are for the non-evolving model; the
dashed curves are for a single burst galaxy formed at $z_f=10$; the dotted 
curves are for a single burst galaxy formed at $z_f=5$.  The curves of the 
right panel represent the redshift evolution of (from top to bottom) the 
$R-K$, $R-H$, $I-H$, and $z'-H$ colors.  It is shown that evolved galaxies at
redshifts $z>1$ may be selected based on $R-K \apg 6$, $R-H \apg 4.5$, $I-H 
\apg 3$, or $z'-H \apg 2$.}

\figcaption{Summaries of the image quality of the $H$-band survey.  The left
panel shows the distribution of the FWHM in arcsec of a mean PSF in each tile.
The right panel shows the distribution of the $5\,\sigma$ $H$-band detection 
threshold in each tile estimated based on mean sky noise over a four arcsecond 
diameter aperture.}

\figcaption{An example of an assembled tile in the CDFS region.  The image is
approximately 13.5 arcmin on a side.  The $5\,\sigma$ point source detection 
limit is $H \approx 20.9$ in a four arcsecond diameter aperture in this tile 
and the PSF is $\approx 0.7$ arcsec FWHM.  The bottom panel shows details of an
$\approx 2\times 1$ arcmin$^2$ region.  To illustrate the depth and contrast of
the image, we have selected two objects of $H=19.4$ (left) and 21.3 
(right), respectively, as indicated by the arrows.}

\figcaption{The associated $1\,\sigma$ noise image of Figure 3.  The image 
dimension is the same as Figure 3.  Darker regions represent noiser areas.  The
composition of 16 stacked images, corresponding to four arrays at four 
pointings of the telescope, is clearly illustrated by the ``square'' pattern in
the image.  The noise level varies between arrays and pointings because of the
differences in array sensitivity and total exposure time.  Black dots indicate 
the locations of bright stars, where the noise level is dominated by object 
flux instead of sky.}

\figcaption{The detection efficiency estimated in one of the CDFS $H$ tiles for
objects of different intrinsic profiles and angular sizes.  As indicated in the
upper-left corner, the solid curve represents a stellar profile; the dotted 
curve represents an $r^{1/4}$ de Vaucouleur profile with a half-light radius of
0.3 arcsec; the long-dashed, short-dashed, and dash-dotted curves represent 
disk profiles with scale radii of 0.15, 0.3 and 0.6 arcsec, respectively.  The
thick curves show the detection efficiencies for a point source profile (solid)
and an extended profile (short-dashed) determined empirically by coadding 
individual object images (see \S\ 4.3 for a detailed description).}

\figcaption{Comparisons of model fluxes and various photometric measurements 
for objects of different brightness and intrinsic surface brightness profiles. 
The first column shows the test results for corrected isophotal magnitudes 
provided by SExtractor, the second column for aperture magnitudes within a 
four arcsecond diameter, and the third column for the ``best'' magnitudes 
estimated by SExtractor.  Points are the median residuals and error bars 
indicate the 16th and 84th percentile of the residuals in 1000 realizations.  
The last point in each panel corresponds to the 90\% completeness limit of the 
image, which varies with object intrinsic surface brightness profile.}

\figcaption{Robustness of using the stellarity index provided in SExtractor to
identify stars.  According to Bertin \& Arnouts (1996), a point source has a 
stellarity index close to one, while an extended object has a stellarity index
close to zero.  The solid curve indicates the fraction of the input stars that
have a stellarity index $\apl 0.5$ as a function of $H$-band magnitude.  The 
dashed curve indicates the fraction of stars that have a stellarity index $\apl
0.8$.  The dotted line indicates the 50\% completeness limit of the test image 
for a point source.}

\figcaption{Stellar fraction in the HDFS and CDFS regions.  Closed circles 
indicate the mean surface density of galaxies observed in the two fields; 
stars indicate the surface density of stars in the HDFS region; open triangles 
indicate the surface density of stars in the CDFS region.}

\figcaption{The $I-H$ color versus $H$-band magnitude diagram for the $H$-band
selected galaxies in the HDFS and CDFS regions.  We show the 95\% completeness 
limit of the $H$-band data in the short dashed line and the $5\,\sigma$ 
detection threshold over a four arcsecond diameter aperture for the $I$-band 
images of the HDFS and CDFS regions in the dot-dashed lines. The blue solid 
curve represents a non-evolving $L_*$ elliptical galaxy; the red curve 
represents a passively evolving $L_*$ galaxy formed at $z=10$; and the green 
curve represents an $L_*$ galaxy constantly forming stars since $z=30$.  The 
dashed blue curve represents a non-evolving $3\,L_*$ elliptical galaxy.  The 
dotted line indicates the $I-H =3$ color selection threshold for red galaxies 
at $z\apg 1$.}

\figcaption{The $V-I$ vs. $I-H$ color--color diagram for objects identified
as stars (crosses) and galaxies (filled points) in the HDFS and CDFS regions. 
Open circles indicate galaxies that are not detected in $V$.  Solid curves 
indicate the mean colors of galaxy of different types as a function of redshift
at redshifts between $z=0$ to 2, from elliptical or S0 (yellow), Sab (green), 
Scd (cyan), to irregular (blue) galaxies.  The dotted curve indicates the 
predicted colors of a passively evolving galaxy formed in a single burst at 
redshift $z_f=10$.  The short and long dashed curves indicate the predicted 
colors of galaxies formed at $z_f=30$ with an exponentially declining star 
formation rate $\exp (-t/\tau)$ for $\tau = 1$ and 2 Gyr, respectively.  The 
dash-dotted line indicates the predicted colors of a galaxy formed at $z_f=30$ 
with a constant star formation rate at all redshifts.  Filled circles along 
each curve indicate the corresponding redshifts, starting from $z=2$ on the 
right and evolving toward lower redshifts in steps of $\Delta z = 0.5$.  The 
dotted line indicates the $I-H=3$ color selection threshold for red galaxies at
$z\apg 1$.}

\figcaption{Differential number counts for the $H$-band selected galaxies in
the HDFS and CDFS regions.  The open symbles are previous measurements 
presented by Yan \etal\ (1998; triangles) and Martini (2001b; squares).  The 
curves are predictions based on various star formation scenarios discussed in 
\S\ 6.2.}

\figcaption{Differential number counts for the $H$-band selected red galaxies 
in the HDFS and CDFS regions.  Points are for galaxies of $I-H \apg 3$ and 
squares are for galaxies of $R-H \apg 5$.  As shown in Figure 11, the solid 
curve indicates the prediction based on a non-evolving scenario and the 
short-dashed curve indicates an exponentially declining star formation scenario
for galaxies formed at $z_f = 30$.  These models are described in \S\ 6.3.}

\figcaption{Cumulative surface densities of the total $H$-band detected 
galaxies (open circles) and galaxies of $I-H \apg 3$ (points) and of $R-H \apg 
5$ (squares) as a function of $H$-band magnitude.  The green squares are 
measurements presented by Daddi \etal\ (2000); the blue triangles are from Yan 
\etal\ (2000); and the red cross is from Thompson \etal\ (1999; red cross).  
The solid lines are the best-fit power laws to the cumulative counts.}

\newpage

\plotone{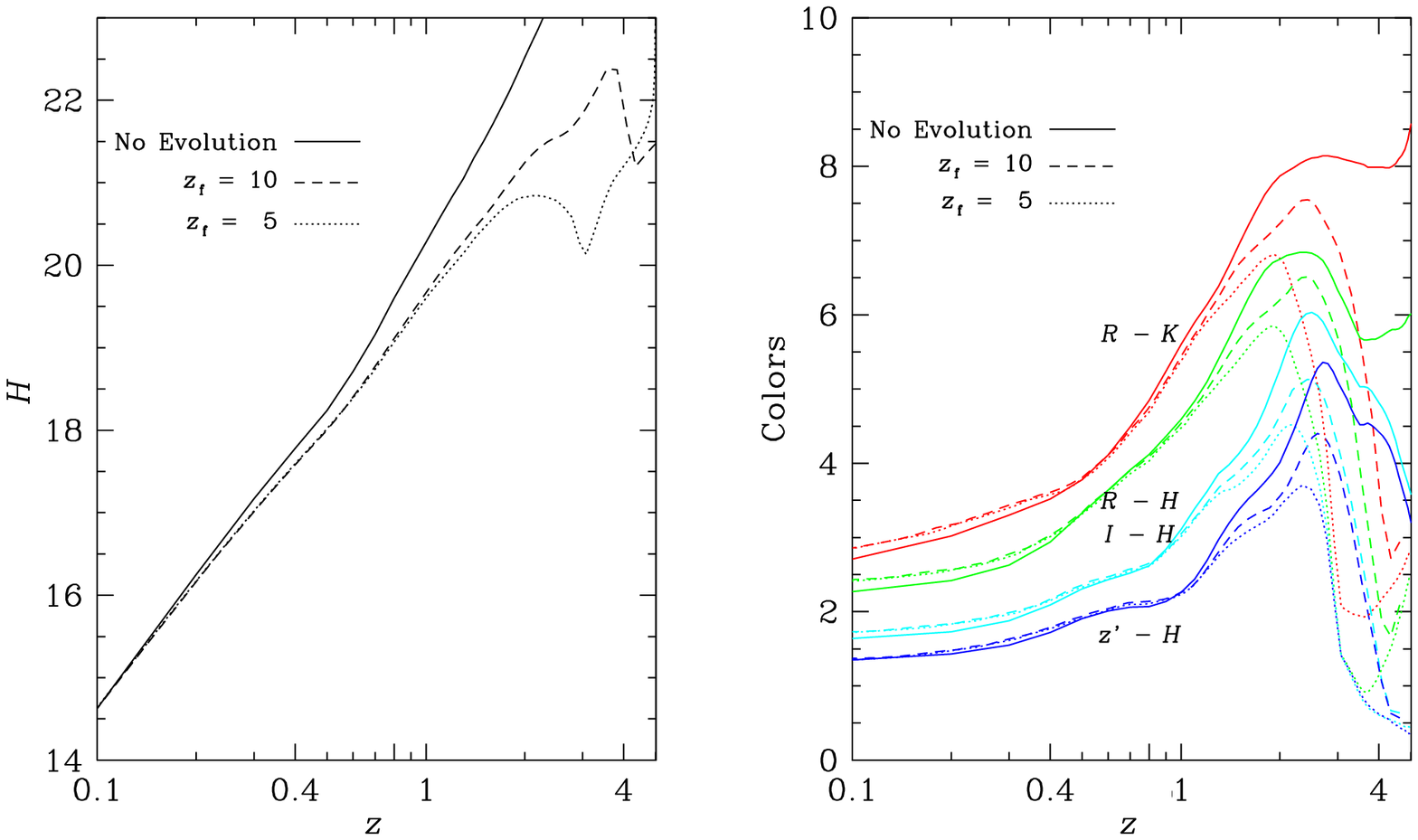}

\newpage

\plotone{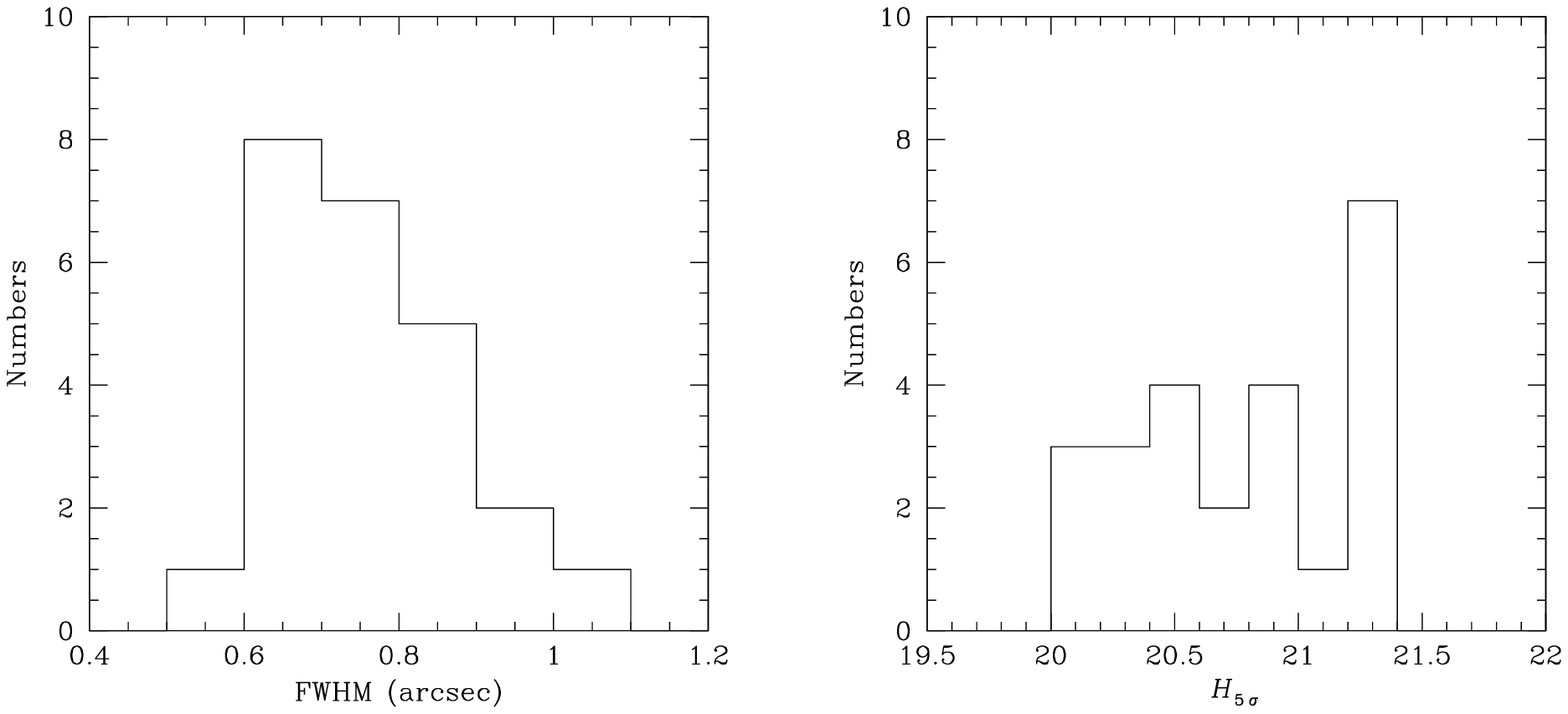}

\newpage


\newpage


\newpage

\plotone{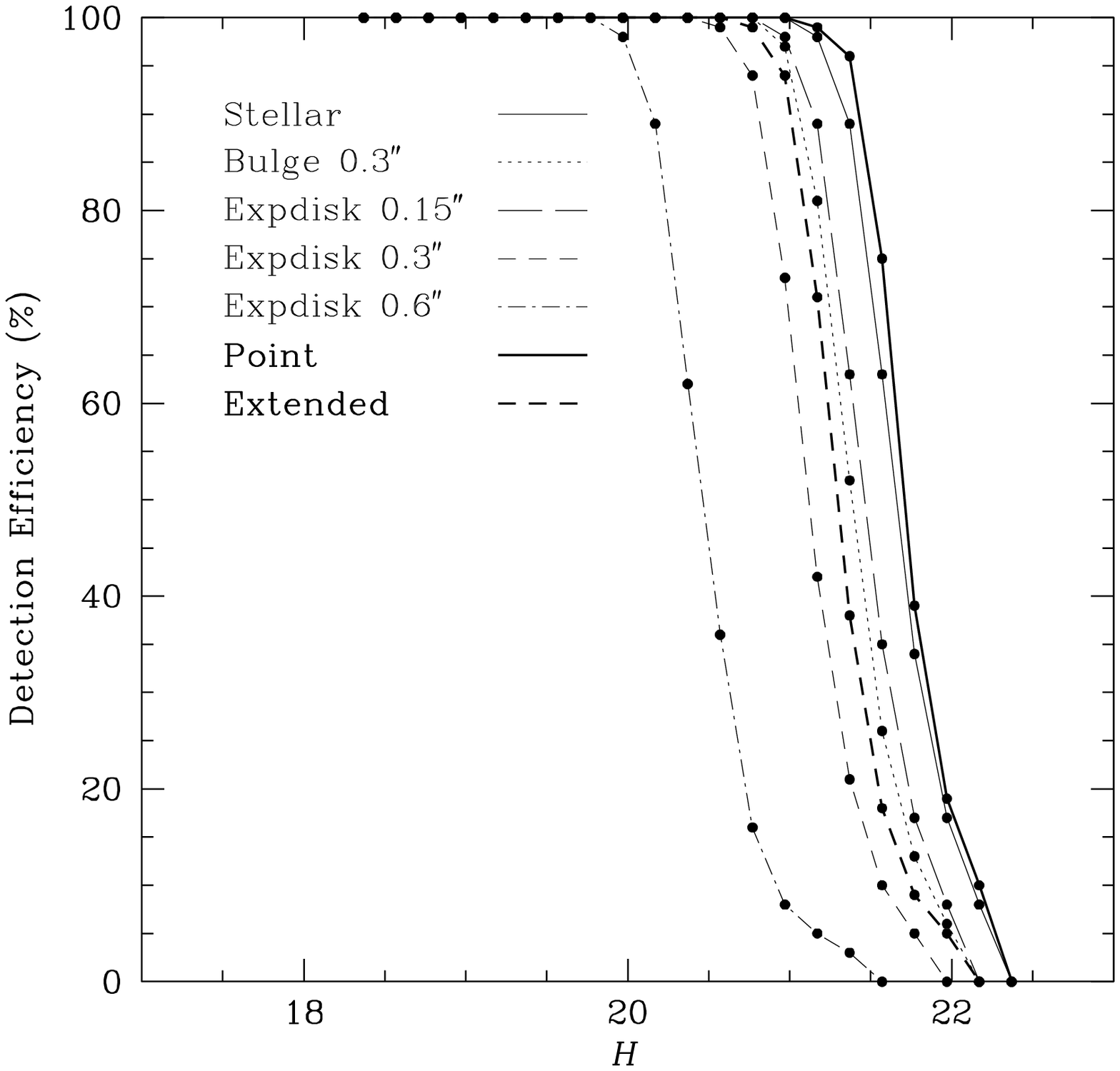}

\newpage

\plotone{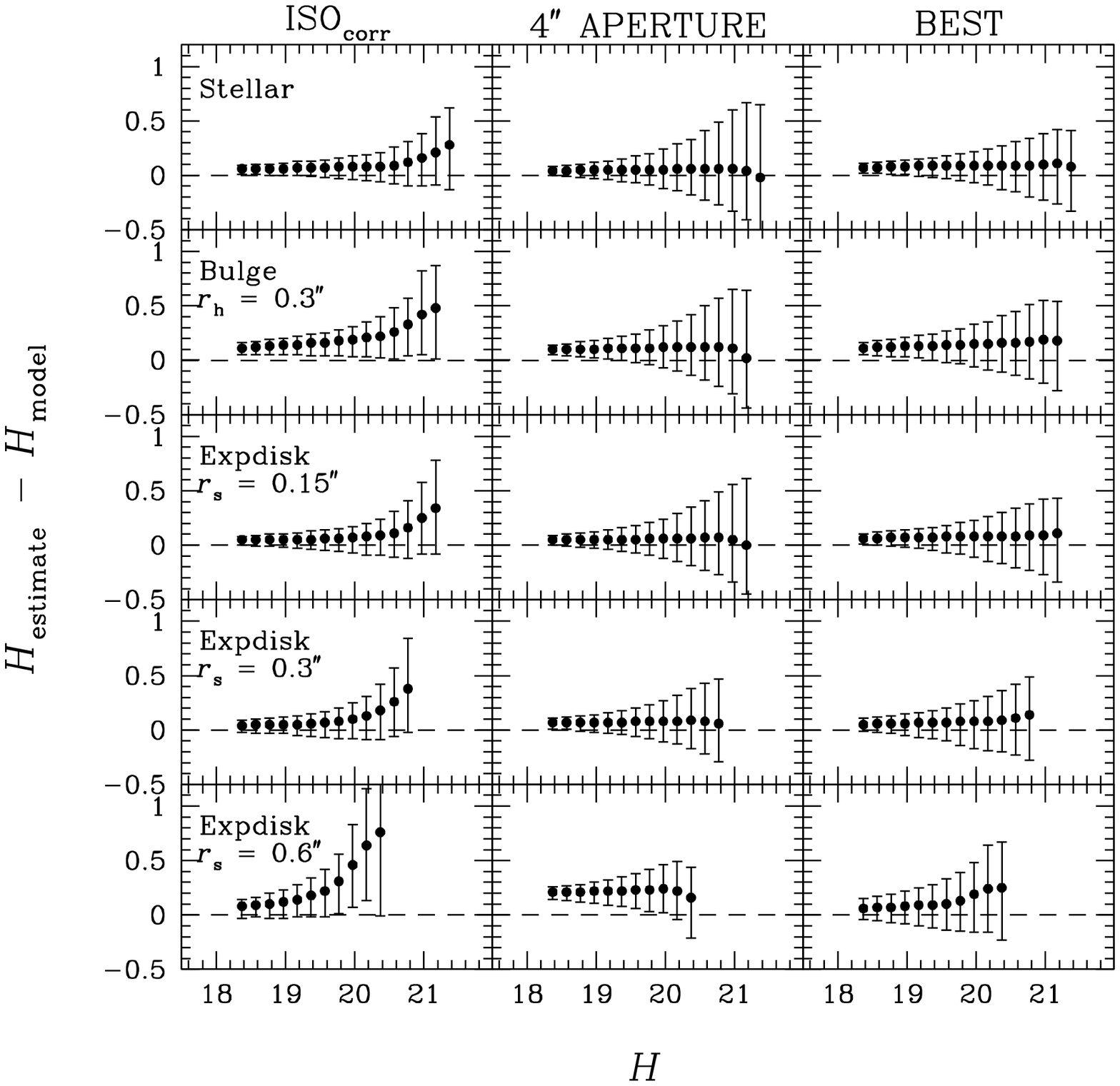}

\newpage

\plotone{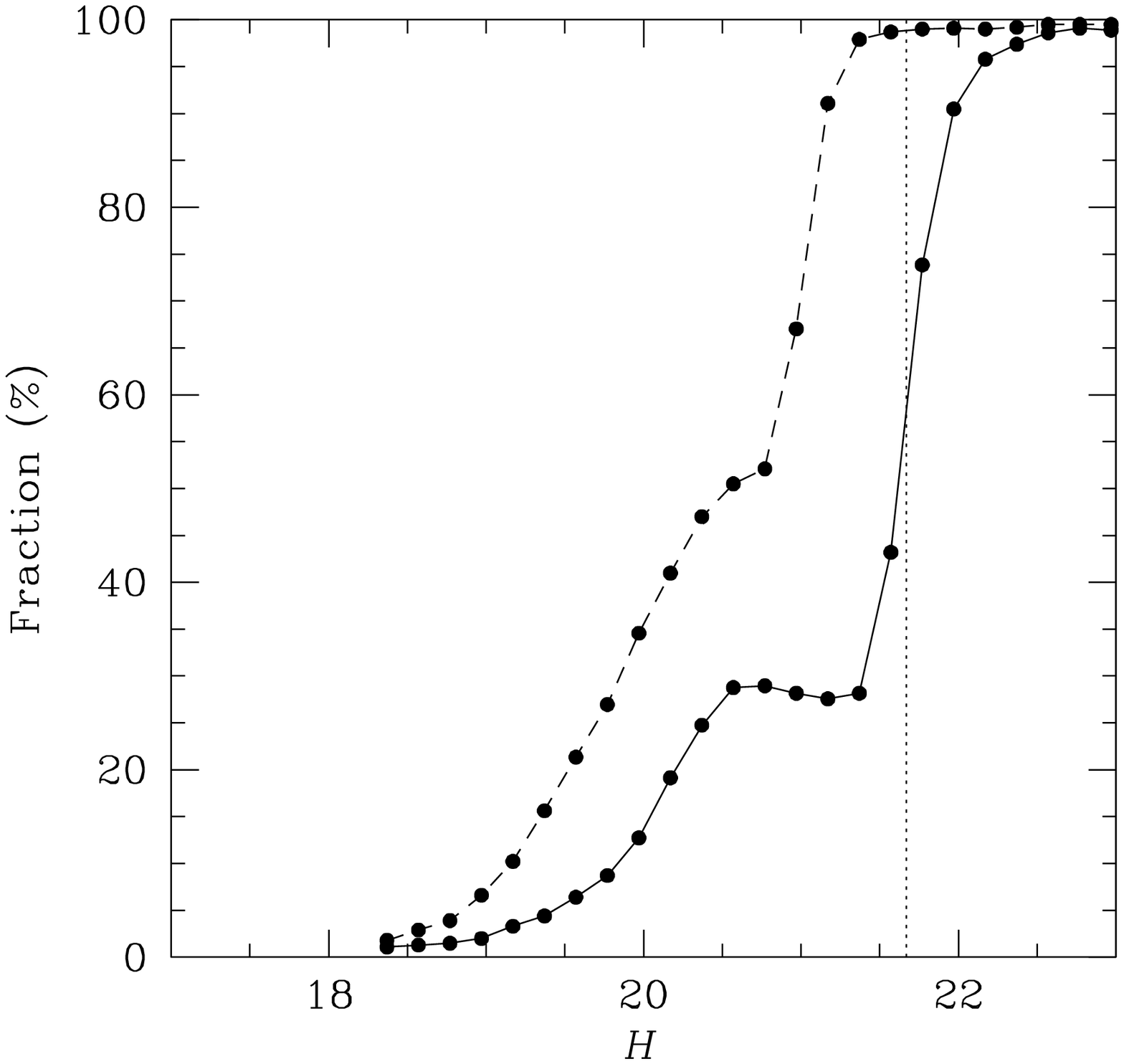}

\newpage

\plotone{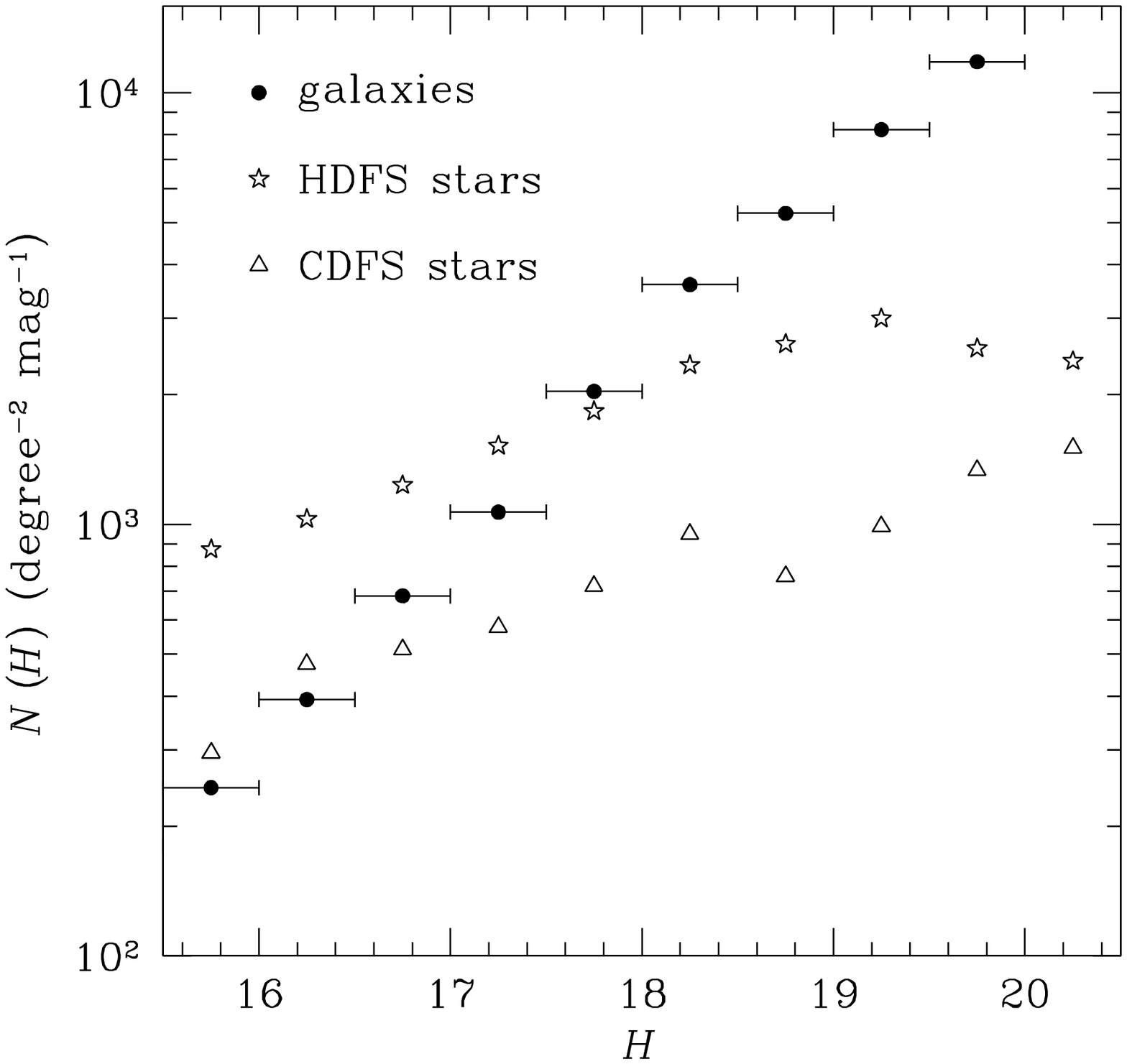}

\newpage

\plotone{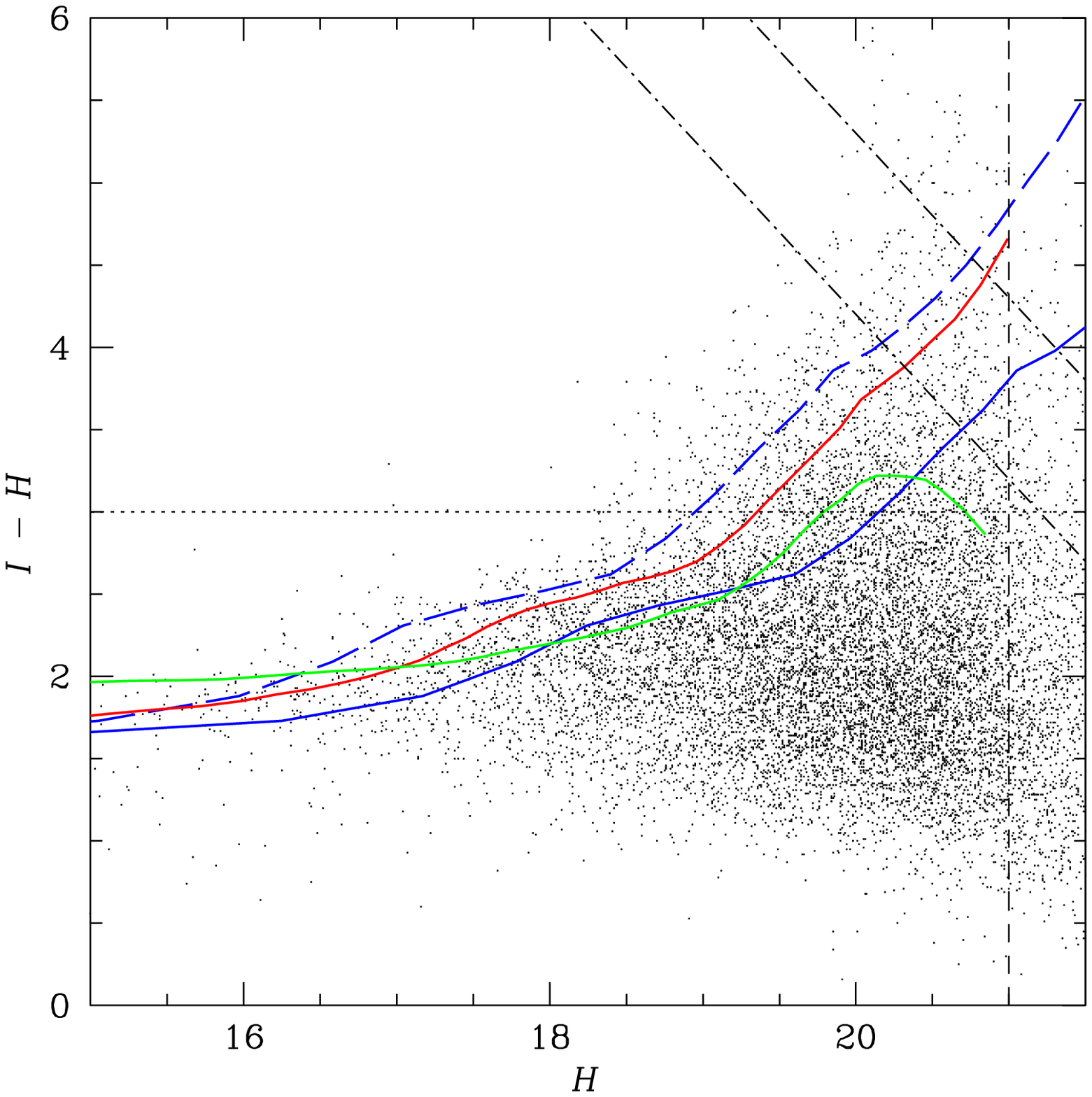}

\newpage


\newpage

\plotone{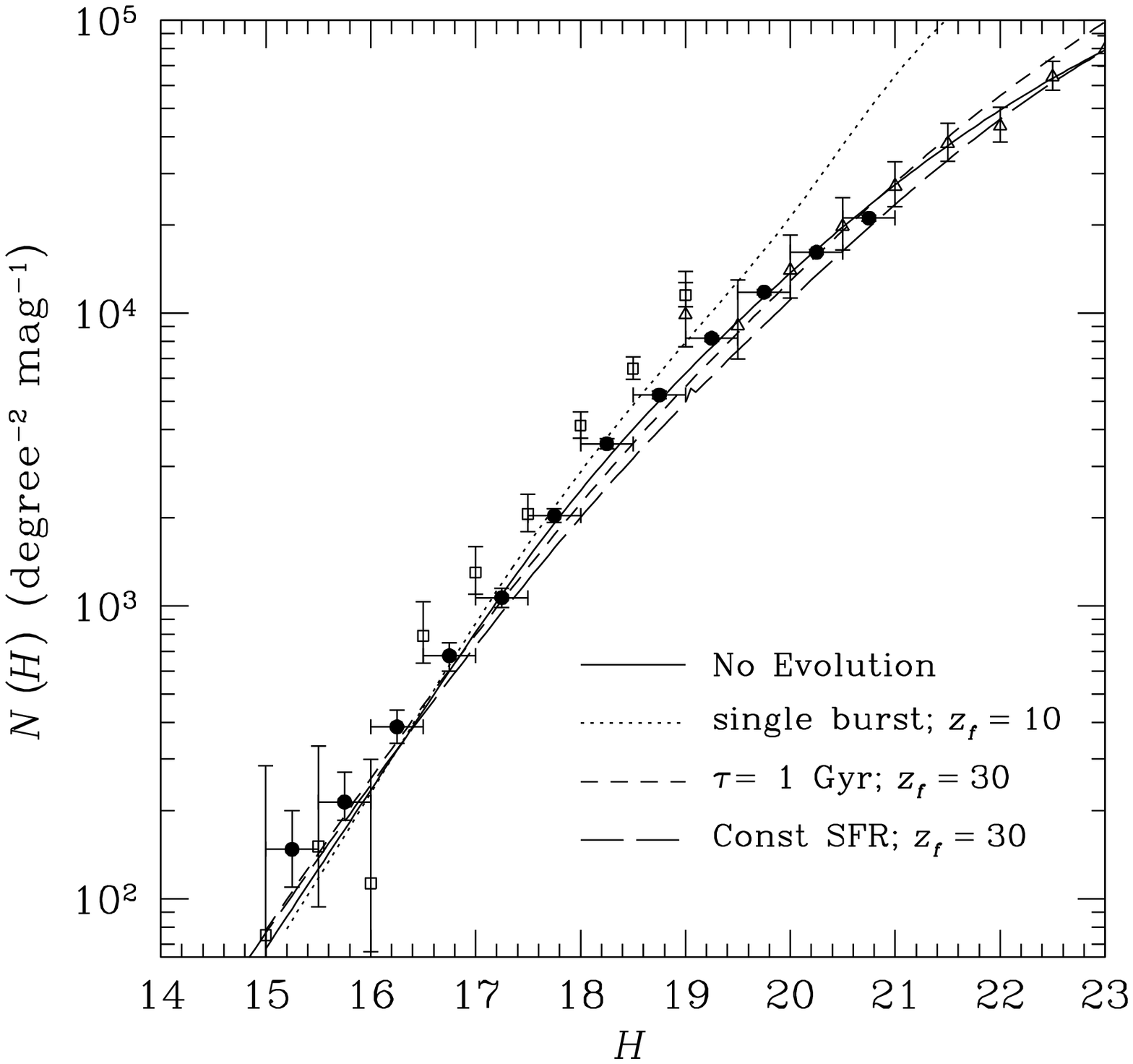}

\newpage

\plotone{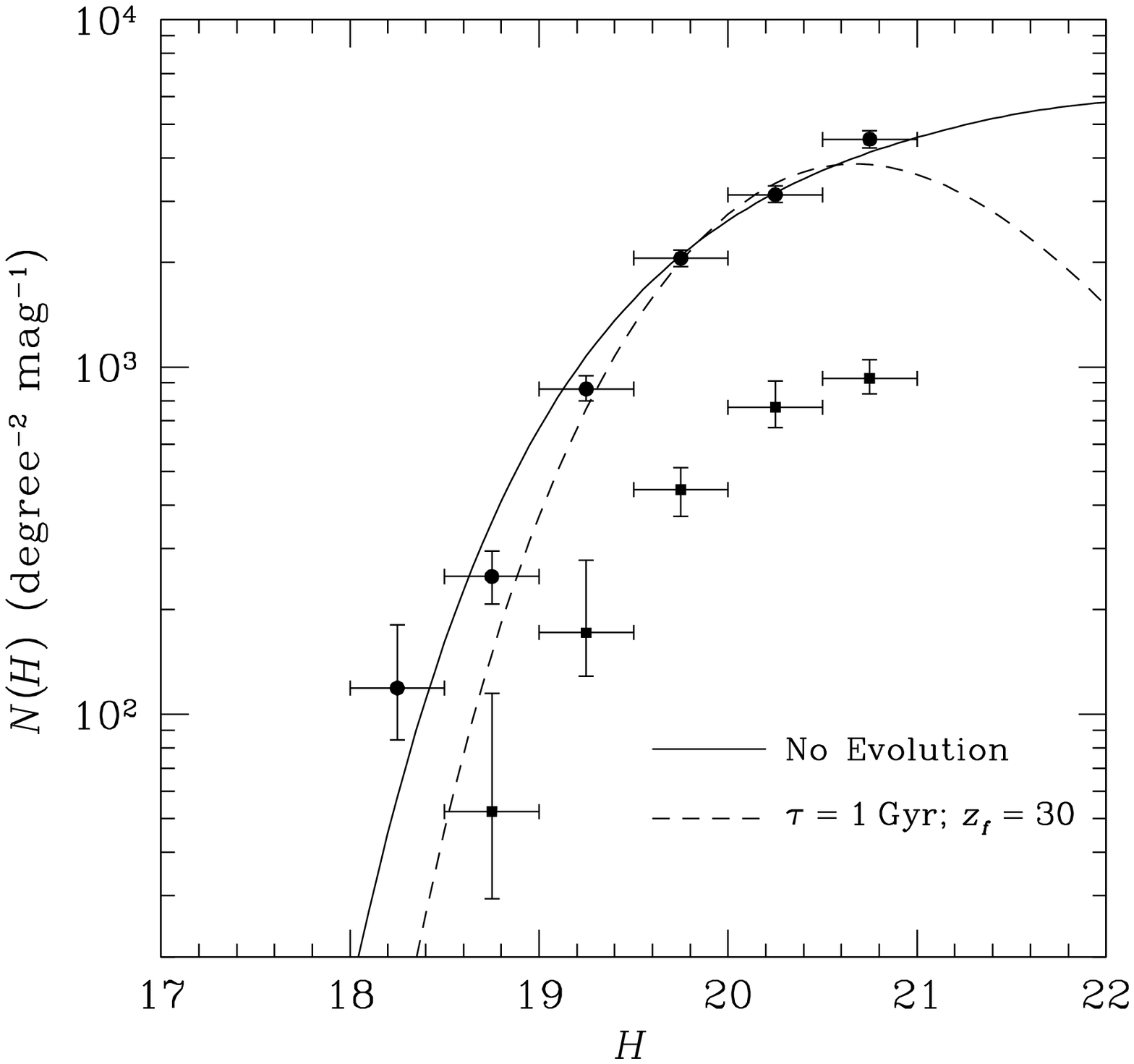}

\newpage

\plotone{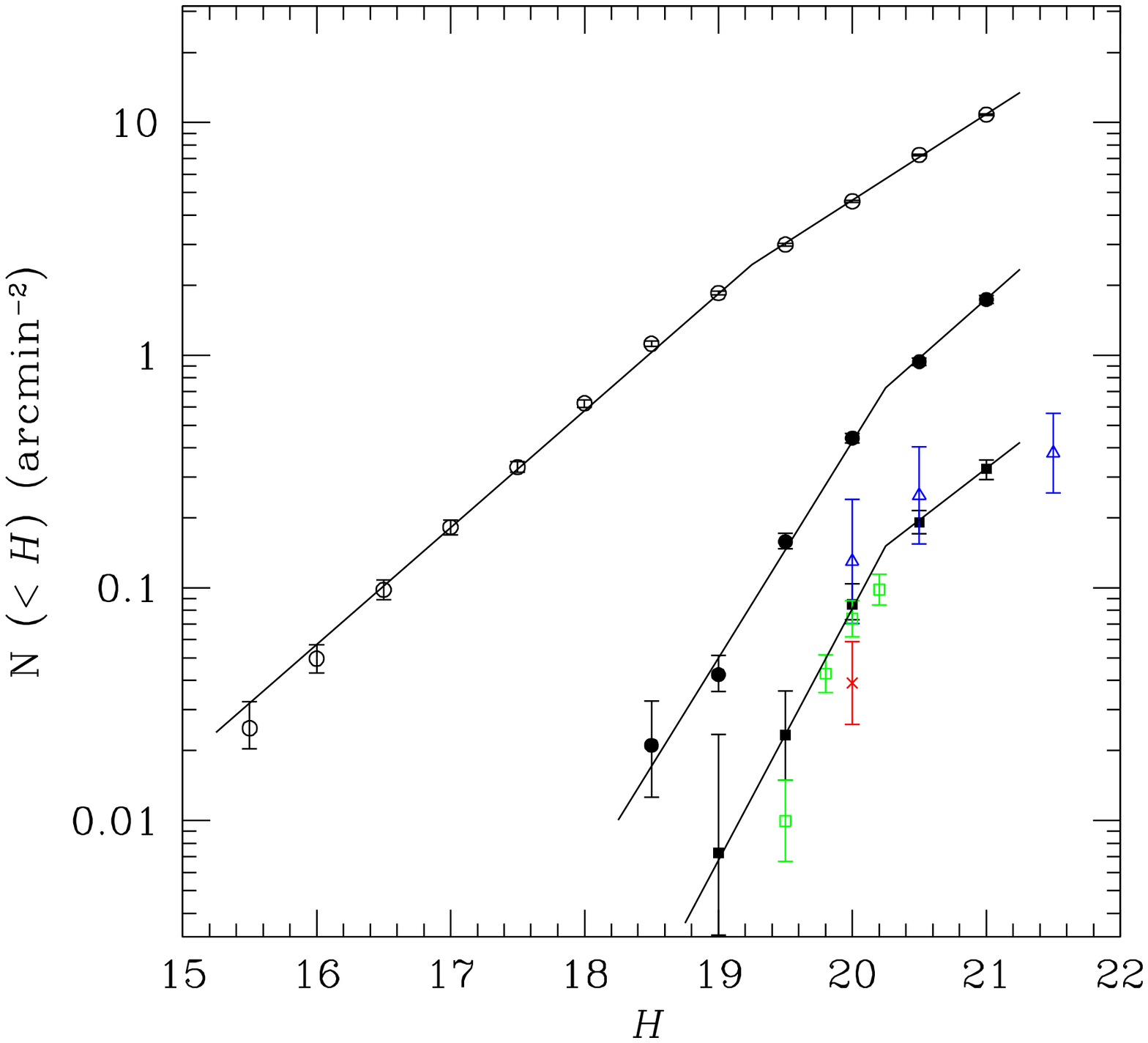}

\end{document}